\documentclass[10pt,conference]{IEEEtran}
\usepackage{amsmath,amssymb,amsfonts}
\usepackage{algorithmic}
\usepackage{graphicx}
\usepackage{textcomp}
\usepackage{multirow}
\usepackage{tabularx}
\usepackage{booktabs}
\usepackage{amsmath}
\usepackage{amsthm}
\usepackage{color}
\usepackage{enumitem}

\usepackage{caption}
\usepackage{subcaption} 
\usepackage{listings}
\usepackage{url}
\usepackage{pgfplots}
\usetikzlibrary{patterns}
\usepackage{color}
\usepackage{cite}
\usepackage{hhline}
\DeclareCaptionFont{black}{\color{black}}
\DeclareCaptionFormat{listing}{\colorbox{gray!50}{\parbox[l]{\columnwidth}{#1#2#3}}}
\usepackage[]{algorithm2e}
\usepackage{algorithmic}

\def\BibTeX{{\rm B\kern-.05em{\sc i\kern-.025em b}\kern-.08emT\kern-.1667em\lower.7ex\hbox{E}\kern-.125emX}}
    
\DeclareCaptionFont{black}{\color{black}}
\DeclareCaptionFormat{listing}{\colorbox{gray!50}{\parbox[l]{\columnwidth}{#1#2#3}}}
\usepackage[]{algorithm2e}
\usepackage{algorithmic}

\newcommand{\ajitha}[1]{\textbf{\color{blue}AR:{#1}}}

\begin{document}

%
% The "title" command has an optional parameter, allowing the author to define a "short title" to be used in page headers.
\title{CAT: Change-focused Android GUI Testing}

%
% The "author" command and its associated commands are used to define the authors and their affiliations.
% Of note is the shared affiliation of the first two authors, and the "authornote" and "authornotemark" commands
% used to denote shared contribution to the research.

\author{\IEEEauthorblockN{Chao Peng}
\IEEEauthorblockA{University of Edinburgh \\
Edinburgh, United Kingdom \\
chao.peng@ed.ac.uk}
\and
\IEEEauthorblockN{Ajitha Rajan}
\IEEEauthorblockA{University of Edinburgh \\
Edinburgh, United Kingdom \\
arajan@ed.ac.uk}
}

%\author{\IEEEauthorblockN{Anonymous Authors}
%	\IEEEauthorblockA{Anonymous Institutes}}

%
% By default, the full list of authors will be used in the page headers. Often, this list is too long, and will overlap
% other information printed in the page headers. This command allows the author to define a more concise list
% of authors' names for this purpose.
%\renewcommand{\shortauthors}{Chao Peng and Ajitha Rajan}

%
% The abstract is a short summary of the work to be presented in the article.

%
% Keywords. The author(s) should pick words that accurately describe the work being
% presented. Separate the keywords with commas.

%
% This command processes the author and affiliation and title information and builds
% the first part of the formatted document.
\maketitle

\begin{abstract}
Android Apps are frequently updated, every couple of weeks, to keep up with changing user, hardware and business demands. 
Correctness of App updates is checked through extensive testing. 
%Mobile Apps are pervading our life and popular Apps on the Google Play App Store are updated frequently.
Recent research has proposed tools for automated GUI event generation in Android Apps. %and regression test selection for Android Apps.
These techniques, however, are not efficient at checking App updates as the generated GUI events do not prioritise updates, and instead explore other App behaviours.  
%However, there is no technique to generate GUI events that targets App updates. % of test case generation for new versions of Apps focusing on the changes. 
%Although regression test input selection is helpful in minimising the number of tests to be performed in the updated version, we still need to write tests specifically for the changed in the source code. 

We address this need in this paper with CAT (Change-focused Android GUI Testing). For App updates, at the source code or GUI level, CAT performs change impact analysis to identify GUI elements affected by the update. 
%Change-impacted GUI elements are highlighted by CAT using its GUI-function mapper and change impact analyser. 
CAT then generates length-3 GUI event sequences to interact with these GUI elements. %, each with the target GUI element being interacted at least once. 

Our empirical evaluations using 21 publicly available open source Android Apps demonstrated that CAT is able to automatically identify GUI elements affected by App updates, generate and execute length-3 GUI event sequences focusing on change-affected GUI elements. Comparison with two popular GUI event generation tools, DroidBot and DroidMate, revealed that CAT was more effective at interacting with the change-affected GUI elements. Finally, CAT was able to detect previously unknown change-related bugs in two  Apps. % related to the changes are revealed by CAT.

\end{abstract}


\section{Introduction}
\label{sec:introduction}
Close to 3 million Apps are available on the Google Play store for Android users.
These Apps are frequently updated, typically every week or two, to keep up with changing user, hardware and business demands. 
To ensure security and correctness, updates in Apps need to be tested thoroughly to ensure changes and 
existing functionality work as expected.

\iffalse
Regression testing [21, 31, 49ś51] is a well-established software
testing technique which ensures that the incremental updates or
the enhancements to the software do not break the existing func-
tionality. There is a multitude of regression testing solutions for
desktop/web applications, but the same cannot be applied directly
to mobile Apps [39] owing to compatibility issues between the for-
mer and the latter’s system architectures. Although developed in
Java, Android runs on the Dalvik virtual machine [14], which is
quite different from the Java virtual machine.
\fi

Several different testing techniques have been proposed in the
literature for testing mobile Apps~\cite{choudhary2015automated,borges2018droidmate, su2017guided,baek2016automated,takala2011experiences, su2016fsmdroid, riganelli2020data, choi2018detreduce, mirzaei2016reducing, song2017ehbdroid}. Majority of existing work focuses on testing only one version
of a mobile App. For updates in Apps, existing test generation work is not effective since it is not focused on updates and may not even exercise them. 
There is existing body of work on regression test selection~\cite{sharma2019qadroid, jiang2018retestdroid, do2016redroid, choi2018detreduce, li2017atom, do2016regression}  - from an existing suite of tests, regression test selection chooses a subset of tests that exercises updates in an App. QADdroid~\cite{sharma2019qadroid} is the only tool in literature that considers changes and their impact at the GUI level when selecting regression tests. Regression test selection techniques only select tests from an existing test suite, they do not generate new tests that exercise changes. None of the existing techniques support GUI test generation targeting updates in Android Apps.  
%
%\ajitha{Did previous studies consider change impact at source code level?}
%\chao{To the best of our knowledge, QADdroid~\cite{sharma2019qadroid} is the only tool in literature analysing change impact of source code to GUI level. It performs control flow graph comparison to check source code differences across two versions and track back to GUI elements. However, it only uses give this information to users for their reference and does not do anything else such as test generation for these events.}

In this paper, we propose a novel approach for generating GUI events targeting App updates. We support updates to source code and the graphical user interface (GUI). 
We design and implement a framework named CAT (Changed-focused Android GUI Testing) that first gathers GUI elements impacted by the update  (referred to as target GUI elements) and then generates GUI events (or test inputs) to exercise these GUI elements. For updates in the source code, CAT's first step of gathering affected GUI elements entails analysing and tracing impact of changes in source code to target GUI elements. To do this, CAT builds a map relating source code functions to GUI elements and associated Android \texttt{Activities} from the App package file (APK). For updates at the GUI level, CAT gathers all the target GUI elements affected by the update and identifies the \texttt{Activities} associated with them. For the next step of GUI event generation, CAT customises an existing model-based android testing tool, DroidBot~\cite{li2017droidbot}, to prioritise event generation for target elements. Additionally, CAT generates length-3 event sequences, rather than single events,  to interact with the target element. Using event sequences allows for more rigorous testing, exercising the target GUI element in different contexts (sequence of events leading to it).    
%Certain app behaviours can only be invoked with a specific order of events. As a result, it is important to test app behaviour with event sequences rather than single events. To address this problem in the context of testing updates, CAT generates length-3 event sequences with one of the events in each sequence interacting with the target element. 

%To do this, CAT first gathers GUI elements and function relations from GUI definitions and function call graphs by parsing the package file (APK) of the subject app. CAT then takes a list of changed functions from the developer as input and map the changes in the GUI-function relation graph to perform change impact analysis and GUI elements that can trigger the changed functions are recorded. Finally, CAT generates GUI event sequences as test input focusing on these GUI elements. Each sequence operates on the target GUI elements at lease once}

We evaluate usefulness and effectiveness of CAT in testing App updates with a dataset of 21  open-source Android Apps from the F-Droid App market. We compare performance of CAT against two state of the art model-based GUI testing tools for Android, DroidBot (DB) and DroidMate (DM)~\cite{li2017droidbot,borges2018droidmate}. For each App, we generate 1000 input GUI events with all three tools. 

CAT was able to trace updates to target GUI elements and generate length-3 event sequences interacting with them in all 21 Apps. We found CAT interacted with the target GUI elements more frequently than DB and DM -- 69 interactions on average for CAT versus 5 for DB and 3 for DM. %\ajitha{these numbers are different from your numbers, check commented line.}
%
%69 more events than DB and 71 more than DM out of 1000 events in total. 
We found events generated by CAT interact with the target GUI elements sooner than other tools owing to CAT's prioritisation of target elements in event generation. %, helped it reach the target element faster than DM by 77 events and similar to DB.
Finally, CAT was able to reveal previously undetected bugs in two Apps in the dataset -- \texttt{World Weather} and \texttt{BeeCount}. DM did not reveal bugs in any Apps, while DB revealed a bug in the \texttt{World Weather} App but not \texttt{BeeCount}. We find the combination of target element priority in event generation along with rigorous target element interaction with length-3 event sequences makes CAT an effective test generation tool for App updates.

In summary, the main contributions in this paper are,  
\begin{enumerate}[topsep=0pt, itemsep=0pt,  partopsep = 0pt]
 \item Novel GUI test generation technique targeting Android App updates, with support for change impact analysis from source code to GUI level. %to identify GUI elements affected by updates in source code.  Affected GUI elements are exercised with length-3 event sequences. 
 \item Empirical evaluation comparing performance of CAT against DB and DM using 21 open-source Android Apps.  
\end{enumerate}

\section{Motivation}
\label{sec:motivation}
In this section, we highlight the need for changed-focused test input generation with a motivating example -- an actively-maintained Android App, Amaze File Manager\footnote{\url{https://github.com/TeamAmaze/AmazeFileManager}}, with more than 3,000 stars on GitHub.
Amaze File Manager is used for Android filesystem management. In addition to basic file operations such as copy and paste, it also supports compression, encryption and cloud service synchronisation. We use the latest version, 3.4.3, of Amaze File Manager in this paper.
Commit history for the project revealed that the \textit{ZipService.java} source file was updated in the latest version. This file contains the implementation for the file compression functionality. 

Changes in \textit{ZipService.java} affect dependent files, \textit{ProcessViewerFragment.java} and \textit{MainActivityHelper.java}. 
These two Java files are linked to the \texttt{compress} option in the main menu, according to the layout file \textit{menu/contextual.xml} used by the \textit{MainActivity} toolbar. Thus, the target GUI element affected by the update is the \texttt{compress} option. We checked this by clicking the \texttt{compress} option, and found it exercised the changed code in \textit{ZipService.java}.

%To understand the impact of updates in \textit{ZipService.java} on other files, we iterate through the source tree of this app, and find  \textit{ProcessViewerFragment.java} and \textit{MainActivityHelper.java} is dependent on \textit{ZipService.java}. 
%Additionally, these two Java files are linked to a layout file used by the \textit{MainActivity} toolbar.
%By going one step further, we find that these two source files are linked to a layout file which is further used by the toolbar of \textit{MainActivity}.

\begin{figure}
	\centering
	\includegraphics[trim = 1cm 0cm 1cm 0cm, scale=0.55]{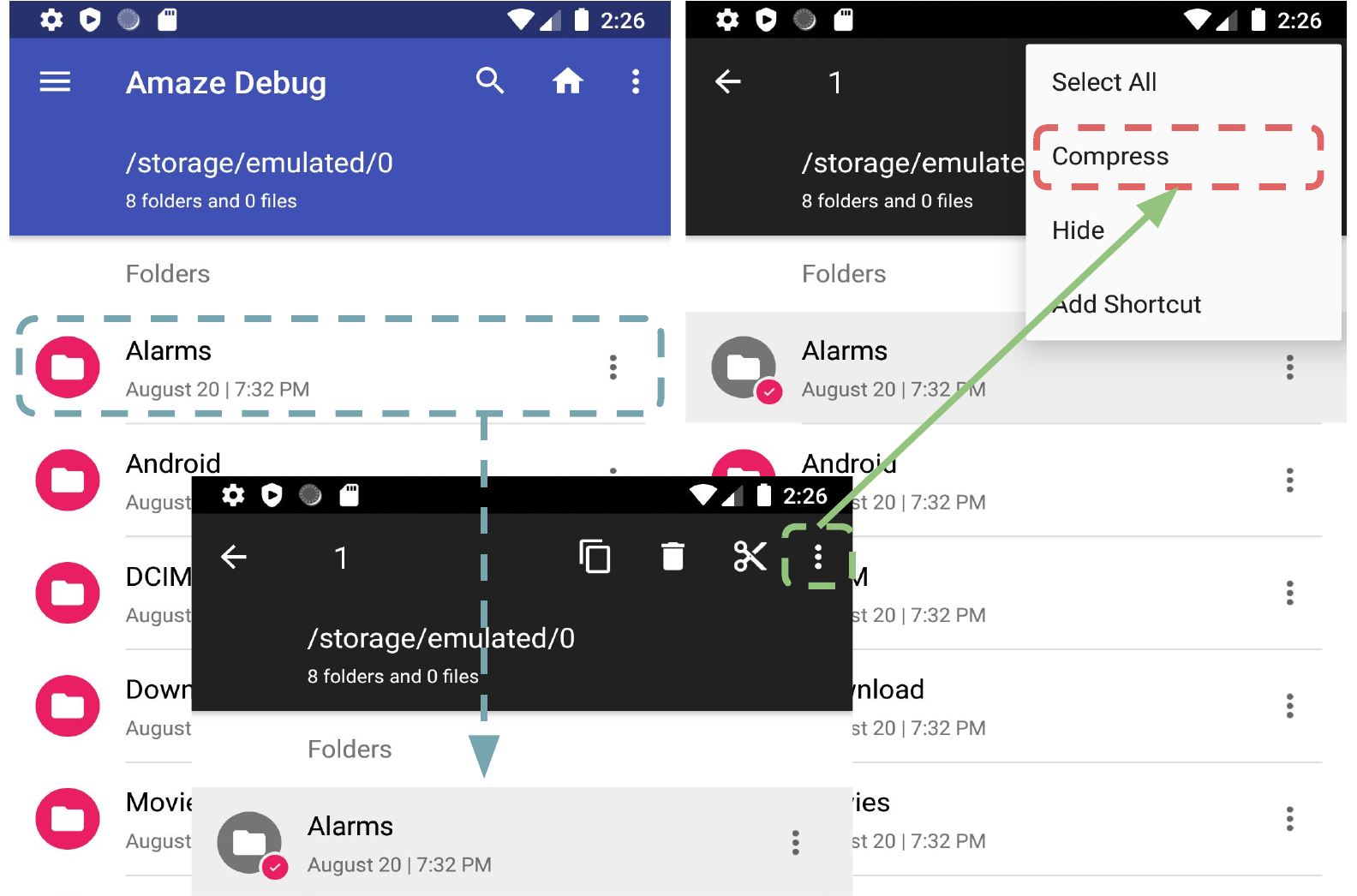}
	\caption{Sequence of input events to interact with \texttt{Compress} option in Amaze File Manager App}
	%\vspace{-12pt}
	\label{fig:example}
\end{figure}

Figure~\ref{fig:example} illustrates the sequence of input events to interact with the target GUI element--\texttt{compress} option: long clicking one item in the file list, clicking the menu icon, and then clicking the \texttt{compress} option.
The \texttt{compress} option is included in the toolbar layout of the main \texttt{Activity} and this layout only appears after long clicking a file or folder. 

When we run existing test input generation tools such as DM~\cite{borges2018droidmate} and DB~\cite{li2017droidbot} on this App, they explore the GUI model without prioritising interactions with the \textit{Compress} option. As a result, these tools may not even exercise the target GUI elements. 
We monitored events generated by DB as an example, and found it entered the final view in Figure 1 after 59 GUI events. %choose a folder in the list followed by clicking the menu button to make the \texttt{compress} option appear after performing 59 GUI events. 
It takes a further 442 GUI events to click the \texttt{compress} option  (as it explores the \texttt{Add ShortCut} option in the menu first) which finally triggers the updated implementation in \textit{ZipService.java}. %Only after 442 GUI events are conducted, during which period this menu is shown 10 times, it is the first time it clicks the \textit{Compress} option and triggers the changed source code. 

The uncertainity observed with existing testing tools in exercising target GUI elements raises the  
need for a GUI test generation tool that prioritises interactions with these GUI elements. We address this need with CAT. 
%However, if we mark this menu as a special state and when this state is matched during exercising the GUI of the app, we ask DB to preferentially play with the \textit{Compress} option, within 1000 events, it enters this menu 13 times and clicks the \textit{Compress} option 6 times.
%In the next sections, we present background on Android app development. %In stead of simply preferentially clicking these target GUI elements, we also add different sequences of related GUI events based on a length-3 GUI coverage model. Details of our approach is discussed in Section~\ref{sec:approach} and the effectiveness of our framework compared to other testing tools is described in Section~\ref{sec:result}.

\section{Background}
\label{sec:background}
Before we discuss our approach in detail, we briefly introduce basic concepts in Android App development and testing.

\subsection{Android Apps}
\label{sec:android-platform}

Android Apps run on top of a stack with three main software layers -- the \texttt{Android application framework}, the \texttt{Android Runtime} and the \texttt{Android-customised Linux kernel}. 
The Android application framework provides high-level services via its API in Java classes so that Android Apps do not need to interact with the operating system directly. 
%So far, there have been 24 publicly available framework releases and consequent changes in the API. 
%Framework versioning is the first element that causes the fragmentation problem in Android. 
%Since it takes several months for a new framework release to become predominant on Android devices, most of the devices in the field run older versions of the framework. 
%Android developers thus constantly have to make an effort to make their apps compatible with older framework versions.

The second layer, the Android runtime provides a Dalvik Virtual Machine which is a kind of Java Virtual Machine but specially designed and optimized for Android. 
A new Android runtime, ART (Android Run-time) is included in recent releases of Android with improved performance and will eventually replace the Dalvik Virtual Machine in the future. 

The customised Linux kernel is the last and bottom layer. This layer plays the role of operating system which provides drivers of hardware such as camera, sensors, the screen and physical keys. 
Low-level memory and network management is also handled by this layer.
On top of the Linux kernel, there is a set of libraries programmed in native code including the browser kernel, video and audio decoding, graphics and internet securities. 

%A set of native code libraries, such as WebKit, libc and SSL, communicate directly with the kernel and provide basic hardware abstractions to the runtime layer.

%At runtime, the Zygote daemon creates a separate Dalvik Virtual Machine (Dalvik VM) for each running app, where a Dalvik VM is a register-based VM that can interpret Dalvik bytecode.
%The most recent version of Android includes radical changes in the runtime layer, as it introduces ART (i.e., Android Run-Time), a new runtime environment that dramatically improves app performance and will eventually replace the Dalvik VM. 
%The custom Linux kernel, which stands at the bottom of the Android software stack, provides the main functionality of the system. 

Android Apps are commonly programmed using Java or Kotlin that are compiled to Java bytecode.
Native code can also be included to boost performance. 
Java bytecode is translated to Dalvik bytecode and stored in a machine executable file in .dex format. 
Android SDK tools bundle Dalvik bytecode, native code (whenever present) along with any data and resource files into an APK, an Android package, which is an archive file with a \texttt{.apk} extension.
The APK file is all that is needed to install the App on Android devices.

\begin{figure}
	\centering
	\includegraphics[trim = 1cm 0cm 1cm 0cm, scale=0.6]{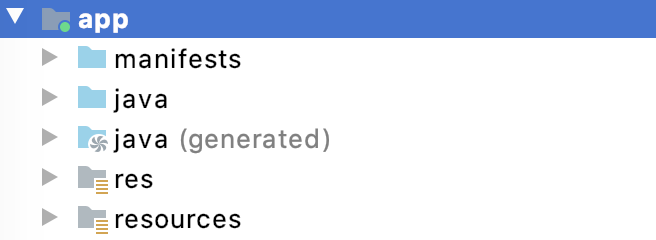}
	\caption{Android project layout}
	%\vspace{-12pt}
	\label{fig:project-layout}
\end{figure}

To build the APK file, an Android project uses the following components: (i) source class files containing source code implementing classes and functions for the App, 
(ii) layout-XML files which defines the GUI layout of all the \texttt{Activities}, 
and (iii) Android manifest which appears in the App root folder as \textit{AndroidManifest.xml} and 
describes essential information about the App -- package name of the App (used to locate the source code), lists of components in the file, user permissions required, hardware and software features used, API libraries needed. Figure~\ref{fig:project-layout} shows the project layout of the Amaze File Manager App in the official IDE, Android Studio~\cite{androidstudio}. The layout contains the mandatory \textit{AndroidManifest.xml} file, class files in the \textit{java} folder and layout, multimedia as well as other resource files in the \textit{res} folder. 
In the rest of this section, we describe terms and concepts in Android App development used in the rest of this paper. 
%Android has project layouts as shown in Figure~\ref{fig:project-layout} according to its documentation of  the official Android IDE (Integrated Development Environment), Android Studio~\cite{androidstudio}. 
%Each project must have a \texttt{AndroidManifest.xml} file which describes essential information about the app -- package name of the app (used to locate the source code), lists of components in the file, user permissions required, hardware and software features used, API libraries needed. 
%Unlike desktop GUI applications, "windows" are referred to as \texttt{activities} in Android development and need to be declared in the \texttt{AndroidManifest.xml} file and implemented in the source code package.

%The folder with the name of the package name contains all source code developed for this app, including GUI interaction functions such as callbacks and listeners. 

%Android apps are mainly written in Java, although it is often the case that developers include native code to improve their performance. 
%Java source code first gets compiled into Java bytecode, then translated into Dalvik bytecode, and finally stored into a machine executable file in .dex format. 
%Apps are finally distributed in the form of apk files, which are compressed folders containing dex files, native code (whenever present), and other application resources. 
%\texttt{Activities} are crucial components in charge of an app’s user interface.
An \texttt{Activity} implements a window or screen in the App containing various \texttt{GUI elements}, such as buttons and text areas. 
Developers can control the behaviour of each \texttt{Activity} by implementing appropriate callbacks for each life-cycle phase (i.e., created, paused, resumed, and destroyed). 
%Activities react to user input events such as clicks, and consequently are the primary target of testing tools for Android.
\texttt{Activities} are first declared in the \textit{AndroidManifest.xml} file and implemented as Java classes in the source code folder. 

\texttt{GUI elements} (also referred to as \texttt{Views} or \texttt{Widgets}) are the basic building blocks for user interactions, such as textboxes, buttons and containers of other \texttt{GUI elements}.
\texttt{Views} can be associated to \texttt{Activities} either in the source code or defined in the XML layout files. An \texttt{Activity} uses a GUI registration function, setContentView(), whose parameter is the identify of a layout file to include \texttt{Views} defined in that layout. 
For instance, the menu as shown in Figure~\ref{fig:example} is a fragment \texttt{Activity} and its menu items are text \texttt{Views}. These \texttt{Views} are defined in the main menu layout XML file and this file is referenced in the menu fragment \texttt{Activity} by calling the setContentView() function.
%In this paper, 
%The GUI layout of each activity is defined in XML files. XML nodes in these files define properties of GUI elements such as style, location and unique identifiers.

\texttt{Views} are responsible for event handling. \texttt{Input events} may be button clicks, edit text, touch, etc. %when the user interacts with GUI elements in the views that defined in a user interface of our application, when the user interacts with it.
To respond to an event of a particular type, the \texttt{View} (or \texttt{GUI element}) must register an appropriate event listener and implement the corresponding callback method (called by the Android Framework when the \texttt{View} is triggered by user interaction).
%bTo handle input events in Android, the views or GUI elements need to have an event listener. %The View class, from which all UI components are derived contains a wide range event listener interfaces and each listener interface contains an abstract declaration for a callback method. To respond to an event of a particular type, the view must register an appropriate event listener and implement the corresponding callback method.
For example, if a button is to respond to a click event it must register to View.onClickListener event listener and implement the corresponding onClick() callback method. When a button click event is detected, the Android framework will call the onClick() method of that particular \texttt{View}.

%We define the following termsto better illustrate our approach and evaluation in the rest of this paper. 
An \texttt{event sequence} is an ordered set of input events. %In the example of Figure~\ref{fig:example}, the event sequence performed to exercise the change code is \{Long Click on a file, Click menu icon, Click Compress\}. 
The term \texttt{state} in this paper refers to GUI state which is a collection of  GUI information about the current screen and all the GUI elements in it. %An \texttt{input event} is able to transit the current state to another. 
Amaze file manager App shown in Figure~\ref{fig:example} has three different states although it remains in the same \texttt{Activity}. 
We refer to changed-affected GUI elements as \texttt{target GUI elements}, Activities containing a target GUI element as \texttt{target Activities} and states containing a target GUI element  as \texttt{target states}. 
%The definition of the target GUI element can be found at the \texttt{target Activity} while at run time, certain states of this \texttt{Activity} may noe contain this GUI element. }

%\subsection{Testing Android}
%\label{sec:testing-android}
%Developers can write test scripts in the \texttt{androidTest} and \texttt{test} sub directories (highlighted in green in Figure~\ref{fig:project-layout}) to run tests on the Android Device and local Java Virtual Machine respectively.
%The \texttt{res} folder is used by developers to store resource files, such as 
%\texttt{layouts} in XML format in which \texttt{views}\footnote{Widgets, or GUI elements such as buttons, textboxes and containers for other GUI elements are referred to as \texttt{views} in Android. } are defined. 
%\texttt{layouts} XML files are assigned to \texttt{activities} classes. \ajitha{Do you need to say this sentence?}
%Resource files may also include pictures, language packages, etc. 
%Finally, Android developers need to choose a build system for their projects. Our motivating example employs a widely-used build system named Gradle, as shown in the last line in the project layout.
%\ajitha{Relevance of this paragraph is not fully clear to me. let's leave it here for now.}

As Android Apps are event-driven, inputs are normally in the
form of  events. % which can either mimic user interactions (UI events), such as clicks, scrolls, and text inputs, or system events, such as the notification of a newly received text message.
%After the general layout of Android projects and key concept of Android GUI design are introduced, we discuss 
%In the rest of this section, how Android projects are tested and why change-focused testing is attractive to Android developers.
%Without third-party testing tools, developers can test their apps in Android Studio using 
%test scripts written in Java or Kotlin,  as mentioned earlier, in the \texttt{androidTest} and \texttt{test} sub directories. 
%Test scripts can be used to test functions and classes by calling and validating their return value or checkign for exceptions. To test GUI elements, 
%GUI interactions can be programmed in these scripts and executed in the device emulator to check their performance. 
%Instead of writing these events manually, developers can also use the Espresso Test Recorder~\cite{espressotest} to record and playback interactions with the GUI. 
%Developers can play with the app in Espresso and the interactions are recorded. 
%After this, the recordings can be used by the emulator in the future to test the GUI automatically.
Writing or recording input events manually can be arduous and time-consuming~\cite{sharma2014quantitative}. %, especially to reveal crashes caused by unusual behaviours. 
Automated input event generation to test Android Apps is an active area of research. A summary of existing research in Android testing is presented in the next section. %During a lifecycle of Android development, developers are not always interested in testing the whole app and are sometimes concerned by the correctness of changes made to the current version. Previous work on change-based Android testing focuses on selecting the subset of existing test scripts based on changes or updating properties of GUI elements in the test script if they are changed~\cite{sharma2019qadroid, jiang2018retestdroid, do2016redroid, choi2018detreduce, li2017atom, do2016regression, chang2018change}. There exists no means of new test input generation based on changes.
%%%%%%%%%%%USE THIS MAYBE?%%%%%%%%%%
\iffalse
Despite being GUI-based and mainly written in Java,
Android Apps significantly differ from Java standalone GUI
applications and manifest somehow different kinds of bugs [14],
[15]. Existing test input generation tools for Java [12], [21],
[22] cannot therefore be straightforwardly used to test Android
Apps, and custom tools must be created instead. For this reason,
a great deal of research has been performed in this area, and
several test input generation techniques and tools for Android
Apps have been proposed. The next section provides an overview
of the main existing tools in this arena
\fi
%%%%%%%%%%%%%%%%%%%%%%%

\section{Related Work}
\label{sec:related}
CAT is the first Android GUI testing work focusing on App updates. %Android apps are event-driven and testing these apps on GUI level uses event sequences as test inputs. 
In this Section, we summarise existing work on Android GUI test generation, split into random and model driven testing. We also discuss related research in regression test selection that selects tests based on App updates. % In addition to test generation techniques, we also compare modules of our framework to existing work with similar domains including change impact analysis and change-based test generation.

%\ajitha{I suggest splitting existing work into - 1. Test generation for Android - Under this say with GUI (random testing, and other techniwues)  and without. 2. Regression test selection - change impact, input selection. Skip call graph constructin, you just have FlowDroid within it. You can just attribute call graph to FlowDroid in the approach section.  }

\paragraph{Android GUI test generation}

\textit{Random Testing.} Android Monkey~\cite{androidmonkey} is a popular random testing tool that examines the GUI and randomly selects events to be exercised in the current state until the number of exercised events exceeds the limit set by user. 
%Instead of construct fully random events where testers have no means of building interested event flows,
%allowing the s constructing paths the tester wants to explore and amplifies it with more unexpected events. 
DynoDroid~\cite{machiry2013dynodroid} uses heuristics to select input events rather than being fully random. However, DynoDroid has not been maintained for years and only supports Android version 2.3.5 (Android 10 is current version).
Wetzlmaier et al.~\cite{wetzlmaier2017hybrid} amplify existing test inputs by injecting random test inputs. This technique gives the user more control than Monkey.
None of the existing random testing tools focus on App updates. 
%However, Monkey generates events for the entire state and all these tools are not tailored for changed impacted GUI elements.

\textit{Model-based Android Testing.} DroidBot (DB)~\cite{li2017droidbot} and DroidMate (DM)~\cite{jamrozik2016droidmate, borges2018droidmate} focus on generating test inputs based on GUI models. DB queries the GUI model of the subject App, computes and executes possible events in this model. %It builds events when is app is running and supports depth-first and breadth-first exploration of GUI states. 
DM guides test generation on-the-fly using the GUI model. %DB uses direct communication with the emulator for API monitoring while DM instruments the app to enable the communication with it. 
DB provides an easy to use interface for App exploration. CAT leverages this feature in DB for depth-first App exploration from the start \texttt{Activity}. % to reach GUI state containing change affected GUI elements. % and use as an executois used as an execution base of other work such as Data Loss Detector~\cite{riganelli2020data} which directs DB to trigger more events that are found to be more fragile to data losses in Android including rotating the screen. With the similar ideology, our work also uses DB as the test executor to trigger GUI elements affected by changes more frequently to test changes.

\paragraph{Regression test selection} Several studies have examined selection of regression tests based on App updates and their impact. 
Focus of CAT is different - test generation for change affected elements. None of the regression test selection work perform test generation. Nevertheless, both CAT and regression test selection techniques rely on change affected elements identified using change impact analysis. We summarise change impact analysis in Android regression test selection below.
%Existing work in this area with needs of locating changes and analysing change impact across versions of subject apps uses different techniques to achieve their goal.
%Chang impact analysis is an important field in software engineering and there has been numerous studies on change impact analysis for desktop, mobile apps as well as other programming domains~\cite{apiwattanapong2005efficient, gethers2012integrated, zhang2012faulttracer, garg2012test}.

Redroid~\cite{do2016redroid, do2016regression} and ReTestDroid~\cite{jiang2018retestdroid} are regression test selection techniques that compare Java source files from original and updated App versions to identify changes and compute change impact at the source code level. Regression tests that exercise change impacted code are selected by the tools. ReTestDroid handles more Java features than Redroid, such as fragments, native code and asynchronous tasks. 
%The information produces by these two modules are used to select test cases for regression testing. ReTestDroid~\cite{jiang2018retestdroid} uses a similar approach to regression test selection but they claims a more precise graph containing additional Android features such as fragments, native code and asynchronous tasks.
Both tools perform change impact analysis at the source code level, not considering GUI elements. CAT performs change impact at the source code and GUI level. %also takes effect of changes on GUI elements into account. analyse how the source code is impacted by version changes while our approach aims to exercise changed code at the GUI level.

QADroid~\cite{sharma2019qadroid} and ATOM~\cite{li2017atom} also perform test selection for regression versions of Apps.
QADroid analyses impact of App updates on code and GUI elements. QADroid, like CAT, builds call graphs based on FlowDroid~\cite{arzt2014flowdroid} and links events to function calls using event-function bindings defined in source code. QADroid does not support change impact analysis for dynamic GUI elements, as it does not support Java reflection. %\chao{For instance, FlowDroid does not support Java reflection which enables creating GUI elements at runtime. It resolves reflective calls only if the passed arguments are string constants, thereby missing all other cases. CAT does not have this limitation as these elements are resolved during test execution. }
ATOM~\cite{li2017atom} builds an event-flow graph for each App version, whose nodes are \texttt{Activities} and edges capturing events causing \texttt{Activity} transitions. It then computes a delta graph using event-flow graphs of the updated and original App versions. Only events existing in the delta graph is selected for regression testing.

\section{Our Approach}
\label{sec:approach}
We present CAT -- Change-focused Android GUI Testing -- framework that provides, 1. Change impact analysis and 2. Test input generation for change impacted GUI elements in Android Apps. Our framework is publicly available at \\
\url{https://github.com/CATAndroidTesting/CAT}.

\begin{figure*}
	\centering
	\includegraphics[trim = 1cm 0cm 1cm 0cm, scale=0.38]{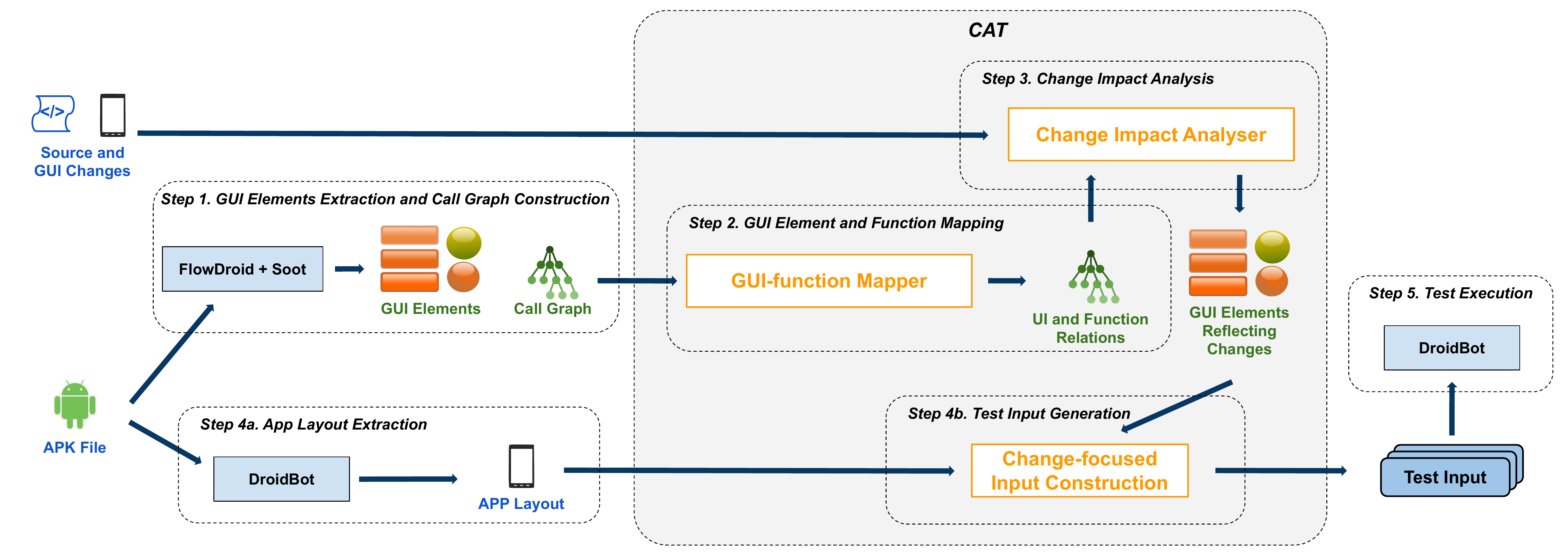}
	\caption{Workflow of the CAT framework.}
	%\vspace{-12pt}
	\label{fig:cat-workflow}
\end{figure*}

The workflow of CAT is presented in Figure~\ref{fig:cat-workflow}.
The input files to CAT are an APK file and a user-provided change-set for the new version of the App in a JSON file. 
The JSON file lists the signatures of changed classes and functions in the source code. %\ajitha{if the change is at the GUI level, does the user provide anything in the JSON file?} 
%\chao{Changes at the GUI level are not only made in layout files but also refleted in \texttt{Activity} classes and listener functions. Therefore CAT does not require the user to provide the delta layout file.}
Output is a set of GUI event sequences to exercise the changes and change impact in the App. 
Steps in CAT's workflow are as follows, 
\begin{description}
	\item[1. Input Preprocessing.] The APK file is preprocessed by our customised version of FlowDroid~\cite{arzt2014flowdroid} and Soot~\cite{vallee2010soot} to produce a list of layouts in the App, Androidmanifest.XML file, and  a call graph representing calling relations between functions in the Java code. 
	\item[2. GUI Element and Function Mapping.] CAT then traces GUI elements to the underlying listener functions in the source code.  Using this tracing information and the call graph from Step 1, CAT generates a combined graph for tracing between source code functions and GUI elements. %whose nodes are functions or GUI elements, edges show calling relations between functions or GUI elements and functions. %with paths from each GUI element to all the functions it calls directly or indirectly.
	\item[3. Change Impact Analysis.] For functions in the source code that are marked as changed, CAT analyses impact of the changes at the GUI level by tracing paths from the changed functions to GUI elements in the combined graph, generated by Step 2.  %Those paths with at least one node related to the changes are highlighted and tracked all the way to the GUI level, 
	The list of GUI elements generated from this tracing represents the target GUI elements affected by the App update. %\texttt{Views} that can trigger corresponding updates in source code.  %Rather than go through the framework to detect associated changes and affected elements on the GUI, 
%	If information on change impacted GUI elements is available to the user then this step can be skipped
%	The user can optionally specify which GUI elements are changed to skip the change impact analysis and proceed to test generation directly.
	\item[4. Test Input Generation.] For target GUI elements from Step 3, CAT generates test inputs as GUI event sequences that interact with them at least once.
	\item[5. Test Execution.] We execute GUI event sequences generated by CAT on a test execution engine that is a customised version of DB~\cite{li2017droidbot}. 
\end{description}

%Based on the artifacts generated by preprocessing, 

We discuss details of the design and implementation of these working phases of CAT in the rest of this section.

\subsection{Step 1: Input Preprocessing}
\label{sec:input-preprocessing}
%As introduced in Section~\ref{sec:related}, 
%FlowDroid has been widely used to construct call graphs for Android apps taking the APK file as input.
%It is based on Soot, a Java bytecode optimisation and instrumentation framework.
%FlowDroid with Soot provides capabilities for querying information from Android Java bytecode and GUI definitions from Android manifests. 
We use FlowDroid~\cite{arzt2014flowdroid} along with Soot~\cite{vallee2010soot} to preprocess APK files to generate the following - (1) a list of layout files, (2) AndroidManifest.XML file, and (3) a call graph whose nodes are functions in the source code and edges represent calling relations.  
%\ajitha{mapping of GUI elements to Activities -- Flowdroid produces a list of layouts and a list of registration functions that link a given activity to a layout. We analyse the layout to identify the GUI elements within them. Consequently, we can then deduce a mapping between Activities and GUI elements associated with them. } with registration functions such as the \texttt{setContentView} function which is used in \texttt{activity} classes to specify which \texttt{layout} they are using \ajitha{clarify this sentence.}, (2) a call graph with nodes representing functions and edges representing call relations. % all function definitions, function calls and their dependencies. \ajitha{is this an undirecred version of the call graph? Nodes are functions and edges represent calling relations?}

\textit{Customisation:} 
In its current form, FlowDroid does not expose \texttt{layout} files embedded within other \texttt{layout} files and consequently the GUI elements defined in them. 
%FlowDroid parses the information on imported GUI elements but does not make the information accessible to users. 
CAT relies on mapping all GUI elements to \texttt{Activities} and functions in the source code for subsequent steps that analyse change impact and generate GUI event sequences. %trace impact of source code changes on GUI elements in Steps 2 and 3.
We, therefore, augment FlowDroid with a data collector that allows us to gather information on embedded layout files. %\emph{all} GUI element and layout information.
%Some of the information extractions are used by FlowDroid as private fileds for their original needs which are not further capsuled to handlable data formats and are not exposed to the outside.
%We customised FlowDroid to provide us with all GUI element information.

\subsection{Step 2: GUI Element and Function Mapping}
\label{sec:gui-element-and-function-mapping}
In this step, we build a mapping from functions in Java code to GUI elements and \texttt{Activities} using the artifacts generated by FlowDroid in Step 1. 
This mapping will be useful in determining the GUI events that will help exercise the changed functions in Java code. 
The mapping is built in two stages - (1) Mapping functions in Java to GUI elements, and (2) Mapping GUI elements to the \texttt{Activity} class. 

For the first stage, we initially take the call graph produced by FlowDroid and extract the underlying undirected graph from it that captures function dependencies. We refer to this undirected call graph as function graph. Next, we identify \texttt{listeners} in the function graph that get triggered when there is an interaction with a corresponding item in the GUI. We then expand the function graph with additional nodes for GUI elements and edges between listener nodes and the GUI element nodes they register an event for. Output of the first stage is the expanded function graph that contains functions and GUI elements as nodes with undirected edges representing calling relationships. 

In the second stage, we start by extracting all \texttt{Activities} from AndroidManifest.xml. For each \texttt{Activity}, we track the \texttt{setContentView()} method  that is used to render the associated layout at the start. We also track the \texttt{inflate()} method, if present, that is used to change the layout after the \texttt{Activity} starts. 
We use these methods to map each \texttt{Activity} to the Layout it is associated with. The Layout file lists all the GUI elements  that will appear to the user for that \texttt{Activity}. 
We use the GUI element listing in the Layout file along with the \texttt{Activity} - Layout mapping to build an association between GUI elements and the \texttt{Activity} they reside in.

The information from the first and second stages can be merged using GUI elements as the key values connecting both. Merged information allows us to trace functions in Java to GUI elements that can trigger them and further to \texttt{Activities} where the user can interact with these GUI elements. We refer to this merged information as \emph{combined GUI-function map}

\begin{figure}
	\centering
	\includegraphics[trim = 1cm 0cm 1cm 0cm, scale=0.3]{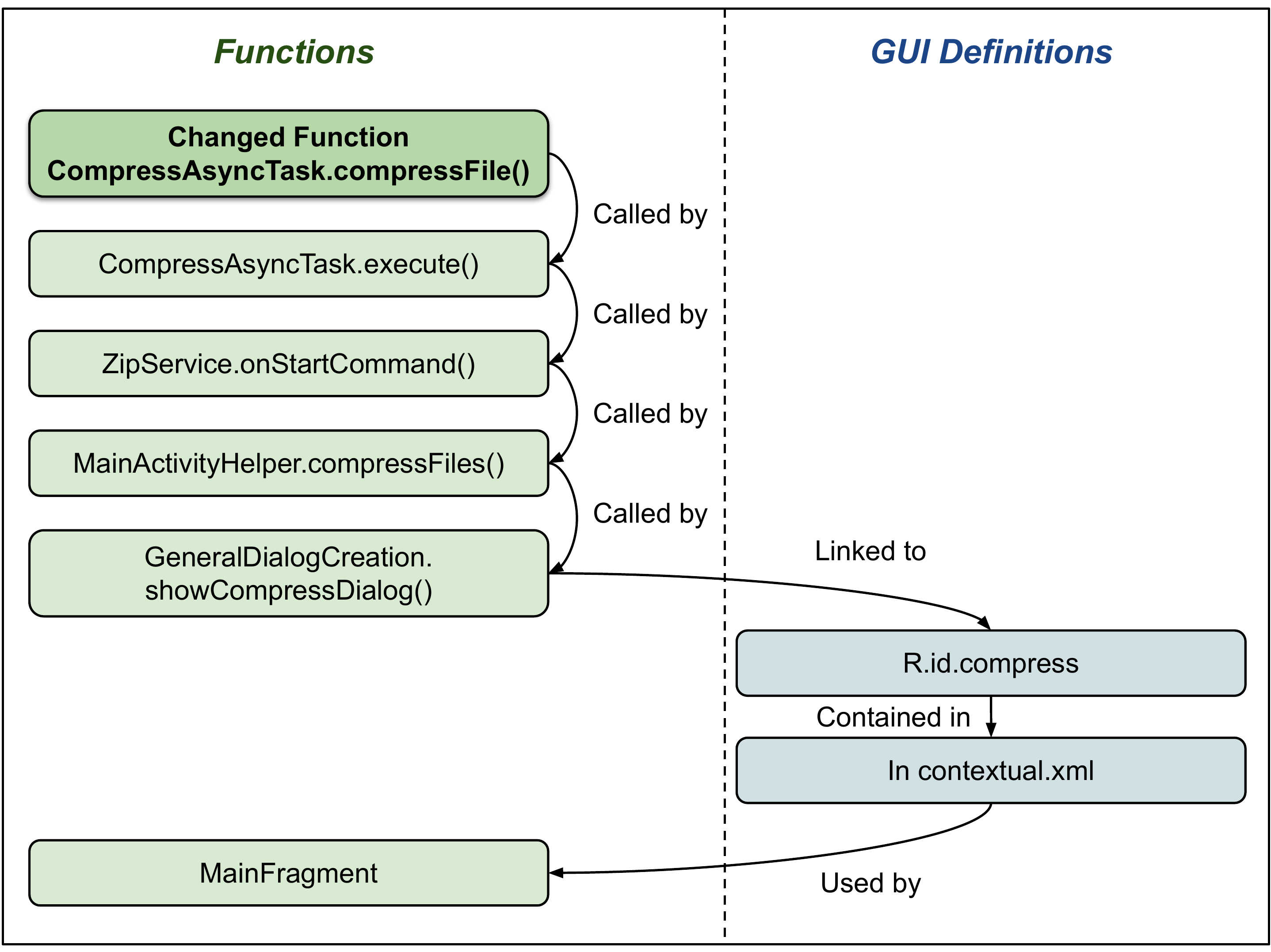}
	\caption{Tracing impact of changed function to GUI element using combined GUI-function map.}
	%\vspace{-12pt}
	\label{fig:graph-example}
\end{figure}

Figure~\ref{fig:graph-example} shows the utility of the combined GUI-function map, built from the Amaze file manager example, to trace from a changed function in the source code to a GUI element and then to an \texttt{Activity} that the GUI elelment is contained in. Dependencies between functions are gathered from the function graph. %Edges between functions and widgets or layouts are built from GUI element registration functions and layout definition files.

Dynamic GUI elements are commonly used in Android development with Java reflection. 
A dynamic GUI element can be referred to in the source code with a symbol whose value is resolved at runtime. %If this GUI element is linked to a change-impacted funciton, we cannot gather information about this GUI element statically.}
To support change impact involving dynamic GUI elements, we annotate these locations so we know to explore them at runtime. When the changed \texttt{Activity} is reached, we record which GUI element leads to this \texttt{Activity}.
It is worth noting that existing static change impact analysis tools, such as QADroid~\cite{sharma2019qadroid}, cannot precisely handle these dynamic features.

\subsection{Step 3: Change Impact Analysis}
\label{sec:change-impact-analysis}

This step starts from the JSON file with signatures of changed functions (\textit{packageName.className.functionName(parameterList)}) identified by the developer. 
%assume the developer provides what changes have been made to this version of the app in the source code level 
We then perform a depth-first traversal of the combined GUI function map starting from the changed functions. We are only interested in visited nodes that are GUI elements for test input generation in the next step. 
%For changed functions, change impact is analysed by performing depth first traversal from the changed nodes in the call graph. 
Transitive closure of all such visited nodes gives the set of target GUI elements.  % which are further aggregated to GUI elements whose listener functions call these impacted functions. 
Events that interact with the target GUI elements are capable of executing the changed function in the source code. 
As shown in Figure~\ref{fig:graph-example}, interacting with the widget with the id \texttt{compress} in the \texttt{MainFragment} \texttt{Activity} can trigger the changed function \texttt{compressFiles()} through a chain of internal function calls.
Output of this step is the set of target GUI elements.

\subsection{Step 4: Test Input Generation}
\label{sec:test-input-generation}

Test input generation with CAT is built on top of DB~\cite{li2017droidbot}. CAT uses DB's depth first exploration from the start \texttt{Activity} to examine different states, checking if the target state (screen with target GUI element) is entered. Once DB enters the target state, CAT takes over event generation, prioritising interactions with target GUI elements in this state. For increased rigor, CAT generates length 3 event sequences, rather than a single event,  to interact with the target GUI element. These steps in input generation are discussed below. 
%DB explores the app under test from the start activity and supports both depth-first and breadth-first search algorithms to seek new states. Before the target state is entered for the first time, CAT uses the depth-first GUI exploration strategy provided by DB to explore as many states as possible. Our contribution in test input generation lies in guiding DB to focus on GUI events capable of triggering source code changes. 

%DB gathers all GUI elements available on the screen and produces a list of all possible events based on their type and property.
%For instance, a \textit{touch event} can be generated for GUI elements that are clickable and a \textit{scroll down event} can be generated for containers with other GUI elements that can be viewed when scrolling down. 
%We accomplish this in two ways. 
%First, we guide DB to enter activities containing change impacted GUI elements, gathered in Step 3. % that are able to exercise changed code.
%Activities associated with these GUI elements can be identified by traversing the combined GUI function graph. 
%This is done by iterating the call graph and locating the list of affected activities.
%Features of these activities are stored in a data file and fed to DB.
\paragraph{Check for the target state} CAT assumes control of input event generation from DB upon first entry into target state. In order to notify CAT, we monitor the states entered by DB, checking if each new state contains a target GUI element identified in Step 3. 
%When DB explores the app, for every new state we check if that state matches features of activities defined in that data file. \ajitha{data file not introduced before}
%DB extracts properties of all GUI elements available in the current state including their ID and class types. 
%CAT checks the GUI element with the ID and class type of the target GUI element is present in the state to perform the match
%For instance, as shown in Figure~\ref{fig:graph-example}, the widget we are interested in is the one which has the ID of \texttt{compress} and is included in the \texttt{contextual} layout. Therefore, the data file in this case contains the \texttt{contextual} layout and every time when DB meets a screen state matching this layout, it means that the screen contains the target widget (\texttt{compress}). 

\paragraph{Prioritise target element interaction} Once we determine that DB has entered the target state, CAT assumes control and generates input events directed at the target GUI element.
%Original test input generated by DB records explores states and steps used to enter these states. If an event cannot enter a new state, this event is not exercised any more. However, our sequence targets the change-impacted GUI elements and is aiming at different orders of playing with these elements are exercised if the budget of running time permits doing so.
In the Amaze file manager App in Figure~\ref{fig:example}, once DB enters the target state containing the \texttt{compress} widget target element, CAT is in control and generates event sequences to  interact with the \texttt{compress} widget. The original DB, in place of CAT, would not prioritise interaction with the target element and would instead treat all GUI elements in the target state as equally interesting. 
%its test input generation policy does not prioritise interacting with the \texttt{compress} widget, instead, it treats all widgets in this state as equally important.  On the other hand, CAT generates event sequences specifically for interacting with the target element. % thereby heavily exercising it. 

\iffalse`
Ideally, testing should try to explore as many orders of event sequences as possible, but it is not feasible in real applications. When the number of interactable widgets increases, the number of all possible permutations of GUI events grows exponentially. However, it is reported that in GUI testing, single event sequences are already able to achieve high statement coverage~\cite{memon2001coverage}. Nearly 95\% statement coverage can be achieved by length-2 event sequences and is slightly further increased to almost 99\% by length-3 event sequence. 
Inspired by these findings, we use the length-3 event sequence coverage as the guidance of our test input generation. 
\fi

\paragraph{Generating length-3 event sequences} %Interacting with the target element can be performed in different contexts, where a context is defined by the sequence of events leading to it. 
For thorough testing of the target element, it would be desirable to interact with it in different contexts, where a context is defined by the sequence of events leading to it. To balance rigor and feasibility, Memon et al.~\cite{memon2001coverage} use Length-3 event sequences in their GUI testing work. Inspired by this, CAT generates length-3 event sequences for interacting with the target element. CAT is the first Android testing tool to consider permutation of events leading to a GUI element interaction.
\iffalse
the screen state containing the changed-impacted GUI elements, CAT gathers all GUI events available on the screen and separates events operating on target GUI elements from events for other elements.
Permutations of length-3 event sequences each with a target GUI element exercised at lease once are computed and generated, and are used to replace the original input sequence generated by DB.
For instance, if four events (\{\textit{A}, \textit{B}, \textit{C}, \textit{D}\}) are available and \textit{Event C} is the target event, CAT first generates all length-2 combinations of \textit{Events A, B, D}, and insert \textit{Event C} to each of these combinations at the beginning, in the middle, and in the end. This approach makes sure that the target event is always exercised in different order with other events. \ajitha{some event sequences may not be feasible. How do you check this?}
\fi
\begin{figure}
	\centering
	\includegraphics[trim = 1cm 0cm 1cm 0cm, scale=0.13]{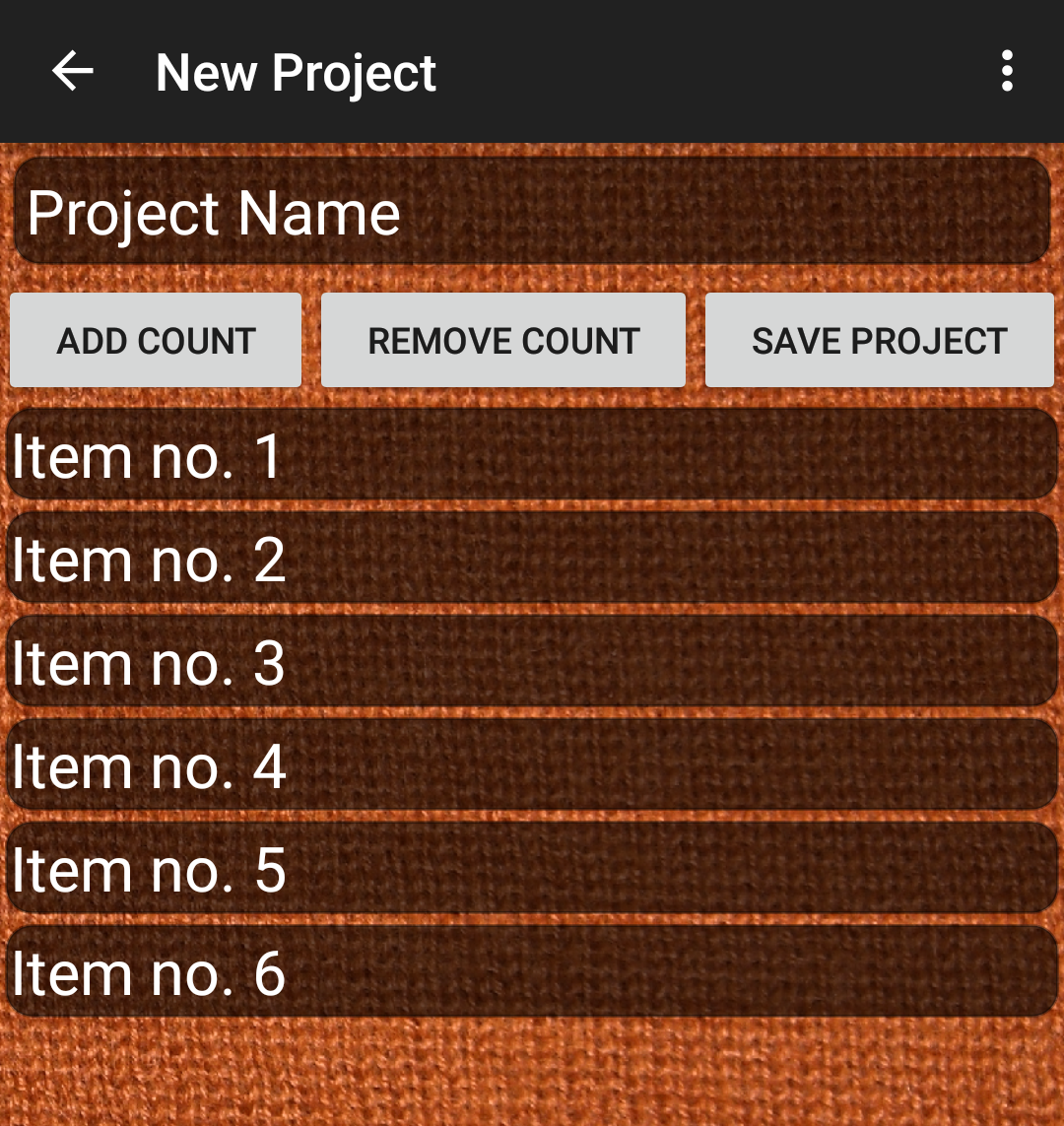}
	\caption{\texttt{New Project} screen in the \texttt{BeeCount} App}
	%\vspace{-12pt}
	\label{fig:bee-count-example}
\end{figure}

We illustrate length-3 event sequences generated by CAT for the \texttt{New Project} screen in the \texttt{BeeCount} App, shown in Figure~\ref{fig:bee-count-example}. Change impact analysis in Step 3 marks the \texttt{SAVE PROJECT} as the target GUI element impacted by changes. %Clicking this button is the target event while editing the textbox and clicking other buttons are other events available.
CAT builds length-3 event sequences that click the \texttt{SAVE PROJECT} option at least once in the sequence. Example sequence generated by CAT is click \texttt{ADD COUNT} -- \texttt{REMOVE COUNT} -- \texttt{SAVE PROJECT}. Sequences like this enable checking the behaviour of \texttt{SAVE PROJECT} using different combinations of prior events. %One of the length-3 event sequences can be \texttt{Input Text}, \texttt{Save Project} and \texttt{Click the Menu Button}. This example is picked from our experiment as reveals a bug which will be further discussed in Section~\ref{sec:bug-findings-result}.

It is worth noting that some generated events can cause the app to leave the target state. When this happens, CAT gives control back to DB which may generate a lot of events exploring states unrelated to updates. To minimise time wasted on other screens, event sequences used to reach every state is recorded by DB and CAT uses these recorded sequences to re-enter the target state. % to go back to the state early. }

\subsection{Step 5: Test Execution}
\label{sec:test-execution}

%Tests are performed by the DB test executor 
Event sequences generated by CAT are executed on the Android Emulator. %when the current activity contains target GUI elements.
%To eliminate redundant test generations and executions, we extended DB so that it only generate tests at the first time when these activities are entered and stored in a data structure. \ajitha{very hard to understand this paragraph.}
For some event sequences, execution of an event may leave the target state, as mentioned earlier. The remaining events in the sequence can only be executed when the App goes back to the target state. 
%one of the events may enter another activity, resulting events in the rest of this sequence cannot be operated in the new activity and should be resumed when going back to the original activity. To make this happen, we implement an extension to DB.
To enable this, CAT gives unique IDs to events and event sequences and marks the execution status of each sequence.
When a sequence has remaining events to be performed but the emulator leaves the target state, the paused events are added to a queue. CAT passes control back to DB to use the recorded steps to go back to target state. When DB returns to the target state, control is passed back to CAT to execute events in the queue, if the queue is non-empty, or other generated event sequences. 
%When the the original activity is back, this queue is queried first to perform remaining operations and when the queue is empty, new event sequences are fetched from the data structure and exercised.

%In our motivating example, test sequences for the \textit{Compression} option is generated at the first time when the menu is shown. For a event sequence, for instance, touch \textit{Add ShortCut}, touching rest of the screen and touching \textit{Compress}, after exercising the first event, the remaining two events are stored in a queue as this event prompts a window for the user to confirm the shortcut information. In this window, other events to be executed are not available. As our changed-focused testing aim is not interested in this window, CAT leaves DB using its default executor to play with this window and waits until our target state is back. When it is back, CAT queries the event queue, fetches the awaiting event, touching rest of the screen, and executes it. This process continues until the queue is empty, after which the next event sequence starts executing. 
%This also results in leaving this screen state, making touching \textit{Compress} unavailable.  This event will be executed next time DB is back to the menu. 

\section{Experiment}
\label{sec:experiment}
\begin{table*}[htbp]
	\caption{Description of subject Apps}
	%\vspace{-12pt}
	\begin{center}
		\begin{tabular}{|l|l|l|l|l|}
		\hline
		\textbf{\#} & \textbf{App Name}    & \textbf{Version} & \textbf{Brief Description}                                   \\ \hline
1           & World Weather        & 1.2.5            & Weather info for cities all over the world                   \\ \hline
2           & Amaze File Manager   & 3.4.3            & Local file manager with cloud synchronisation                \\ \hline
3           & BeeCount             & 2.4.6            & Knitting project helper                                      \\ \hline
4           & Diary                & 1.7.0            & Writing diaries                                              \\ \hline
5           & Omni Notes           & 6.0.5            & Taking notes                                                 \\ \hline
6           & OpenTasks            & 1.2.2            & To do list management                                        \\ \hline
7           & Simple Draw          & 6.1.0            & A simple picture drawing App                                 \\ \hline
8           & Simple File Maganer  & 6.7.3            & Local file management                                        \\ \hline
9           & Simple Solitare      & 3.13             & A collection of card games                                   \\ \hline
10          & WiFiAnalyzer         & 3.0.1            & Wifi network information query                               \\ \hline
11          & Hibi                 & 1.4.0            & Learning Japanese by writing journals                        \\ \hline
12          & Geological Timescale & 0.4.1            & Ecyclopedia of  geological timescale for the earth           \\ \hline
13          & DroidShows           & 7.11.1           & Subscribe and manage favourite TV shows                      \\ \hline
14          & Suntimes             & 0.12.9           & Tracks sunlight and moonlight times                          \\ \hline
15          & Word Scribe          & 1.6.2            & Help story writers list fictional contents in their writings \\ \hline
16          & Nani                 & 0.3.0            & Japanese dictionary                                          \\ \hline
17          & Fate Sheets          & 1.2              & Fate character sheet management                              \\ \hline
18          & Lift                 & 0.2              & Workout and exercise logging                                 \\ \hline
19          & Currency             & 1.33             & Currency rates from the European Central Bank                \\ \hline
20          & Tricky Tripper       & 1.6.2            & Travel expense management                                    \\ \hline
21          & Open Money Box       & 3.4.1            & Budget management                                            \\ \hline
	\end{tabular}
	\label{table:apps}
	\end{center}
	%\vspace{-12pt}
\end{table*}

We evaluate the feasibility and effectiveness of CAT in generating GUI events that exercise target GUI elements. We use 21 Android applications from the F-Droid App market~\footnote{\url{https://f-droid.org/}} that has a catalogue of free and open source Android applications. We investigate the following research requestions:

\begin{description}[topsep = 0pt, itemsep = 0pt]
	\item[Q1. Change Impact Analysis:] 
\textit{Is CAT able to perform change impact analysis from changed functions in code to GUI elements?} %\ajitha{Mention changes at GUI level?}
	%	\textit{In this research question we assess whether CAT is first able to build the combined GUI-function map that contains both the function graph and GUI-function mapping for all apps in our dataset. We then check for changed functions in each app version, if CAT is able to perform change impact analysis to identify GUI elements that can trigger these changes }
	
	To answer this question, we use CAT to first build the combined GUI-function map that contains both the function graph and GUI-function mapping for all Apps in our dataset. We then traverse the combined map from the changed functions to GUI level to identify target elements. We manually verify if the GUI elements identified are correct and complete by going through the source code for each subject App.
%	our implementation to analyse Android applications from our dataset with changes of these apps specified and check if the GUI elements reported by CAT are correct and complete by manually going through the project source code. 
	
	\item[Q2. Test generation for changes:] 
	\textit{Can CAT generate GUI events to exercise target GUI elements faster and more rigorously than popular testing tools -- DroidBot (DB), DroidMate-2 (DM)?}% In terms of time used to reach GUI elements that can trigger changes and statement coverage achieved for the changed source code, how well does CAT perform compared with existing widely-used tools?}
	
	For each subject Android application, we run CAT, DB, DM to generate 1000 tests, where each test is a GUI event. We compare how quickly the tools start interacting with the target GUI element. % by measuring the number of GUI events taken by each tool to reach and interact with the target GUI elements. 
	We measure rigor in exercising target GUI elements as number of generated events that interact with the target elements.  % how many of these events generated by these tools are those events associated with target GUI elements.
	 
	\item[Q3. Bug Finding:] 
	\textit{Is CAT able to detect previously undetected bugs in our App dataset?}
	In this question, we assess if CAT is able to identify previously undetected bugs in our dataset of 21 Android Apps. We compare CAT against DB and DM in this assessment. %its ability to identify previously unknown bugs in our dataset of apps. 
	%We also check if output of CAT is easily usable and can be used to replay bugs. 
	
%	Most importantly, we look forward to seeing bugs identified by CAT and check if the output of CAT is convenient for the developer to replay the bugs. 
\end{description}

\textit{Selected Tools.} We select popular model-based test input generation tools DB and DM for our comparison since these were reported in the literature as being easy-to-use and providing high function code coverage. 
%Features of these tools have been introduced in Section~\ref{sec:related}.
Monkey, a popular random testing tool, is not used in our comparison as it only reports screen coordinates of GUI interactions, not providing details of the interacted GUI element. Mapping the coordinate information to target GUI elements is not trivial, making comparison with our tool difficult.  %Lacking this information, we cannot report the numbers of GUI interactions with the target element for the result analysis.
	%Firstly, we found that this tool generates events without considering state changes after each event execution. When an event leading to the home screen or other screens that do not belong to the subject app, it does not try to go back to the app but remains execution. Most of the events it performs are not within the target app. \ajitha{have you verified this? Given how popular Monkey is, this statement is hard ot believe. }

\textit{Subject Apps.}
We collect subject Apps from the F-Droid App market with released APK files and well maintained commit information. 
Apps available on F-Droid are open source allowing us to perform change impact analysis. % assess how CAT works for Research Question 1.
Table~\ref{table:apps} lists the names and versions of Android Apps used in our experiment. For each of the App versions in our dataset, we manually collect changes in the App by reading commit comments for the APK file release.  A JSON file with signatures of changed functions is input to CAT for change impact analysis.  

Our experiment is performed on an Android emulator running on Mac OS 10.15.6 with 16 GB memory and 2.6 GHz Quad-Core Intel Core i7 processor. The virtual device in the emulator is Google Pixel 2 with Android API 27. The dataset of 21 apps and scripts need to replicate the experiments are available at \url{https://github.com/CATAndroidTesting/CAT}.

\section{Results}
\label{sec:result}
We generated 1000 GUI events with DB, DM and CAT for each of the subject Apps presented in Section~\ref{sec:experiment}. We assess effectiveness of CAT in performing change impact analysis and compare DB, DM and CAT with respect to their effectiveness in exercising target GUI elements. We also report bugs revealed by CAT as a result of the changes. We present our results in the context of the questions in Section~\ref{sec:experiment}.

\subsection{Q1. Change Impact Analysis}
\label{sec:change-impact-analysis-result}
CAT is able to analyse impact of changes to GUI elements automatically for all 21 subject Apps. Change impact analysis could determine target GUI elements affected for 18 of the 21 Apps, statically. Remaining 3 Apps, (\texttt{Tricky Tripper}, \texttt{Suntimes} and \texttt{Hibi}), allocate dynamic GUI elements which required CAT to perform runtime analysis to determine the target GUI elements (described in Section~\ref{sec:change-impact-analysis}). 
%Of the 20 subject apps, CAT is able to analyse change impact at the GUI level for 17 of them automatically at this stage and the remaining 3 (\texttt{Tricky Tripper}, \texttt{Suntimes} and \texttt{Hibi}) have changed activities that do not have explicit link to GUI elements that are able to trigger them. 
%At this step, CAT generates the list of change-impacted GUI elements for those 17 apps. 
%As described in Section~\ref{sec:change-impact-analysis}, for this 3 apps, CAT marks the changed activities at this step and learns which GUI elements are able to trigger them by running the app with a depth-first greedy exploration strategy. 
CAT is able to complete change impact analysis for all the subject Apps within 30 seconds.
Changes in our dataset of Android Apps took different forms. We briefly discuss change impact analysis performed by CAT for these different change types. 
%We summarise different scenarios of changes made to Android apps in our dataset and how they influence change impact analysis of CAT as following:

\begin{description}
    \item[1. New GUI element added.] This is the most straight forward scenario as the information on affected GUI elements is readily available. CAT locates the new GUI element in the layout XML file, marks the GUI element as target element. It then locates the \texttt{Activity} associated with the layout of the target GUI element, and marks it as target \texttt{Activity}. %CAT simply marks the layout with added GUI elements as target state and the added ones as target GUI elements. 
    This scenario appears in 4 subject Apps i.e. \texttt{Simple Draw}, \texttt{World Scribe}, \texttt{Fate Sheets} and \texttt{Nani}.
    \item[2. Modification to existing \texttt{Activity}.] %If an activity is changed, we want to enter this activity to test the behaviour of this activity. 
    The target GUI element in this case is the one that is able to enter the modified \texttt{Activity}. GUI elements implemented to render another \texttt{Activity} may be statically or dynamically allocated. 10 Apps in our dataset had this type of change, with 7 of them implementing static GUI elements to render the modified \texttt{Activity}, and the remaining 3 (\texttt{Tricky Tripper}, \texttt{Suntimes} and \texttt{Hibi}) with dynamically allocated GUI elements. CAT traces changes in the \texttt{Activity} source code to static target GUI elements for 7 Apps. Dynamic target elements in the other 3 Apps are identified during depth-first App exploration. 
\iffalse    
    make the decision with a condition or switch and the decision is made at runtime. For those whose changed \texttt{Activities} is bound to a fixed GUI element, CAT is able to automatically locate these GUI elements in 7 Apps (\texttt{Currency}, \texttt{Geological Times}, \texttt{Wifi Analyzer}, \texttt{Simple Solitare}, \texttt{Diary}, \texttt{BeeCount} and \texttt{World Weather}). The 3 Apps with dynamic GUI elements needs to be explored . To help CAT generate tests for them, we studies the commit history of these Apps and manually marks which GUI elements are able to enter these \texttt{Activities}. \ajitha{I think you should manually mark the changed \texttt{Activities} if that is the ony change and input it to CAT.}
    \fi
    \item[3. Changes to Java functions.] Starting from each changed function, CAT traverses the combined GUI-function map to retrieve the GUI element(s) and associated state and \texttt{Activity} impacted by the change. There were 7 Apps in our dataset with changes to functions and CAT was able to retrieve change affected GUI elements for all 7 Apps. % with an explicit link to a GUI element, either by GUI element listener registration functions or statements of this function which call other functions that changes the behaviour of a GUI element, CAT is able to retrieve this relation directly from the GUI-function graph it builds. The rest of the subject apps falls in this scenario and can also be solved automatically.
\end{description}

\subsection{Q2. Test Generation for Changes}
\label{sec:test-generation-result}

%We generate 1000 GUI events for each app with each of the 3 tools - DM, DB and CAT. 
We assess and compare effectiveness of the tools in exercising the target GUI elements with respect to (1) how quickly they start interaction, and (2) number of target GUI element interactions.  
%The effectiveness of GUI events generated by each of the 3 tools - DM, DB and CAT is assessed by comparing within 1000 events each tools generates, after how many events the target GUI element is out of the total number of events performed in the target state, how many events are interactions with the target GUI element. 
To account for non-determinism in the Android environment, we ran each tool 10 times for each App, generating 1000 GUI events each time. Numbers reported in this section are averaged over the 10 runs.% Table~\ref{table:summary} summarises results achieved by the three tools for each of the 21 subject apps. %The column, \texttt{\#Events before Target Element Interaction}, hsows the efficiency of the tools in reaching the Target GUI element. 

\begin{figure*}[!htbp]
	\centering
	\setlength{\tabcolsep}{1em}
	\setlength{\extrarowheight}{20pt}
	\begin{tabular}{ccc}
		\includegraphics[trim = 0cm 0cm 0cm 0cm, scale=0.5]{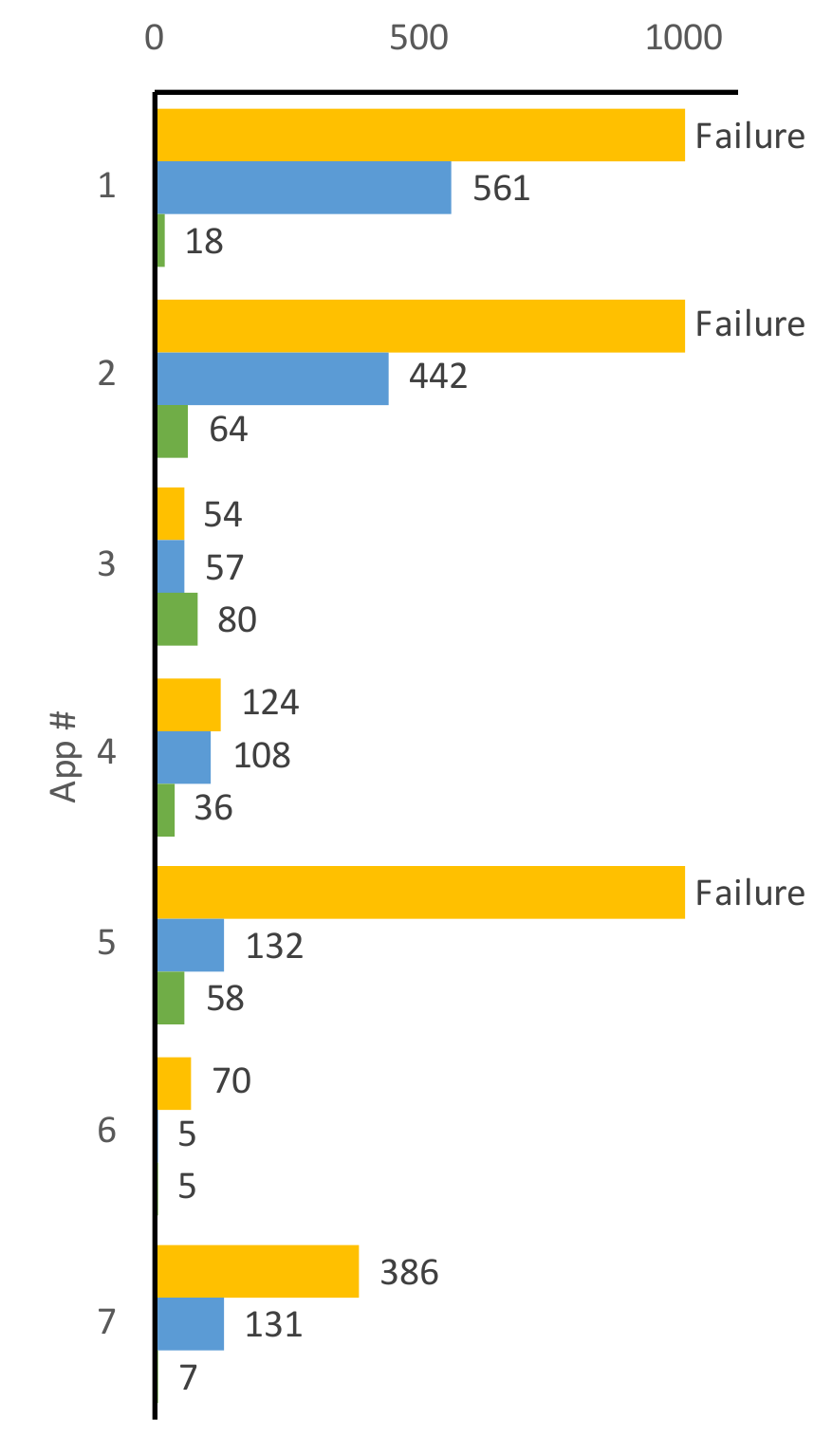}
		&
		\includegraphics[trim = 0cm 0cm 0cm 0cm,scale=0.5]{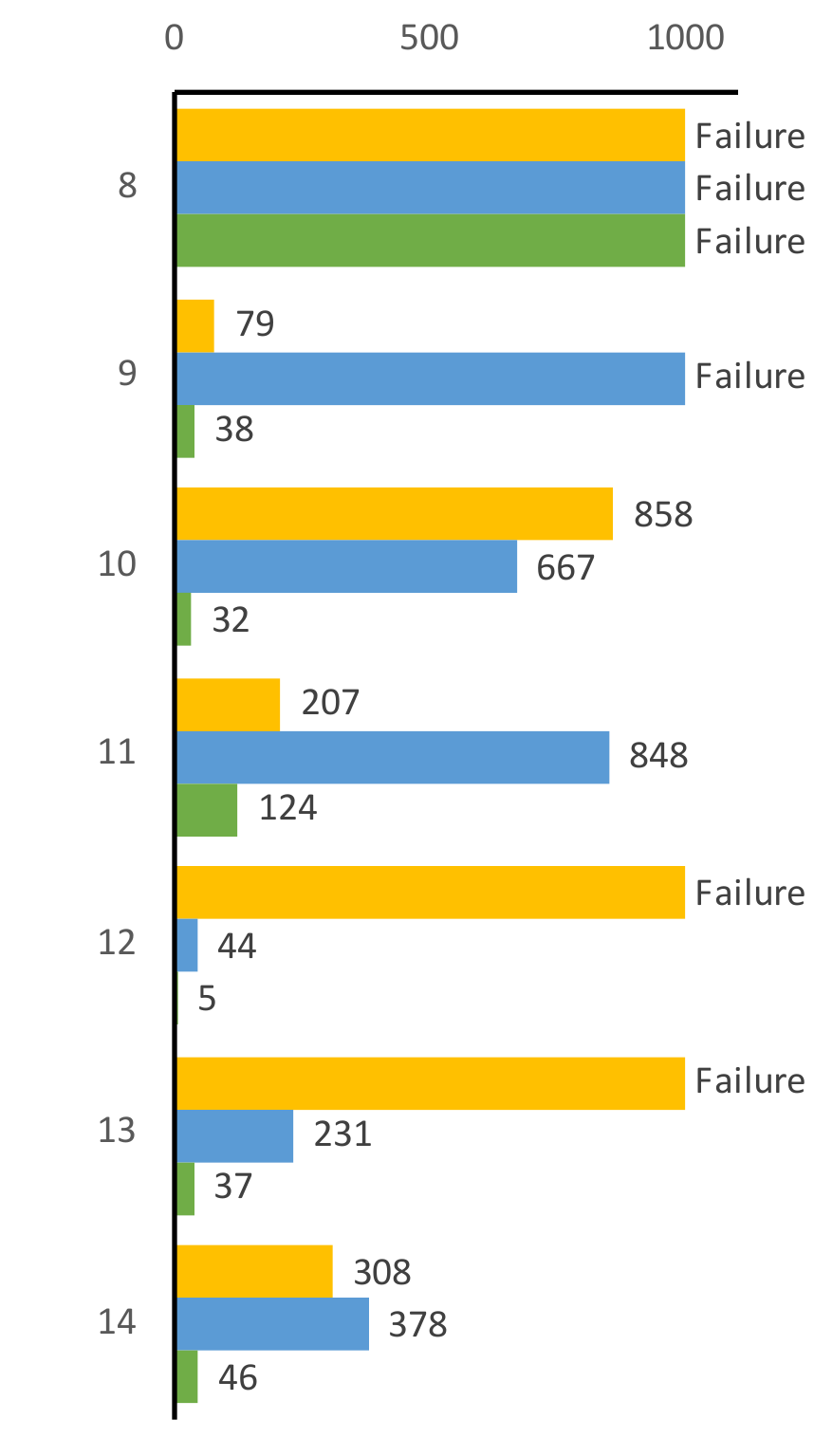}
	&
		\includegraphics[trim = 0cm 0cm 0cm 0cm,scale=0.5]{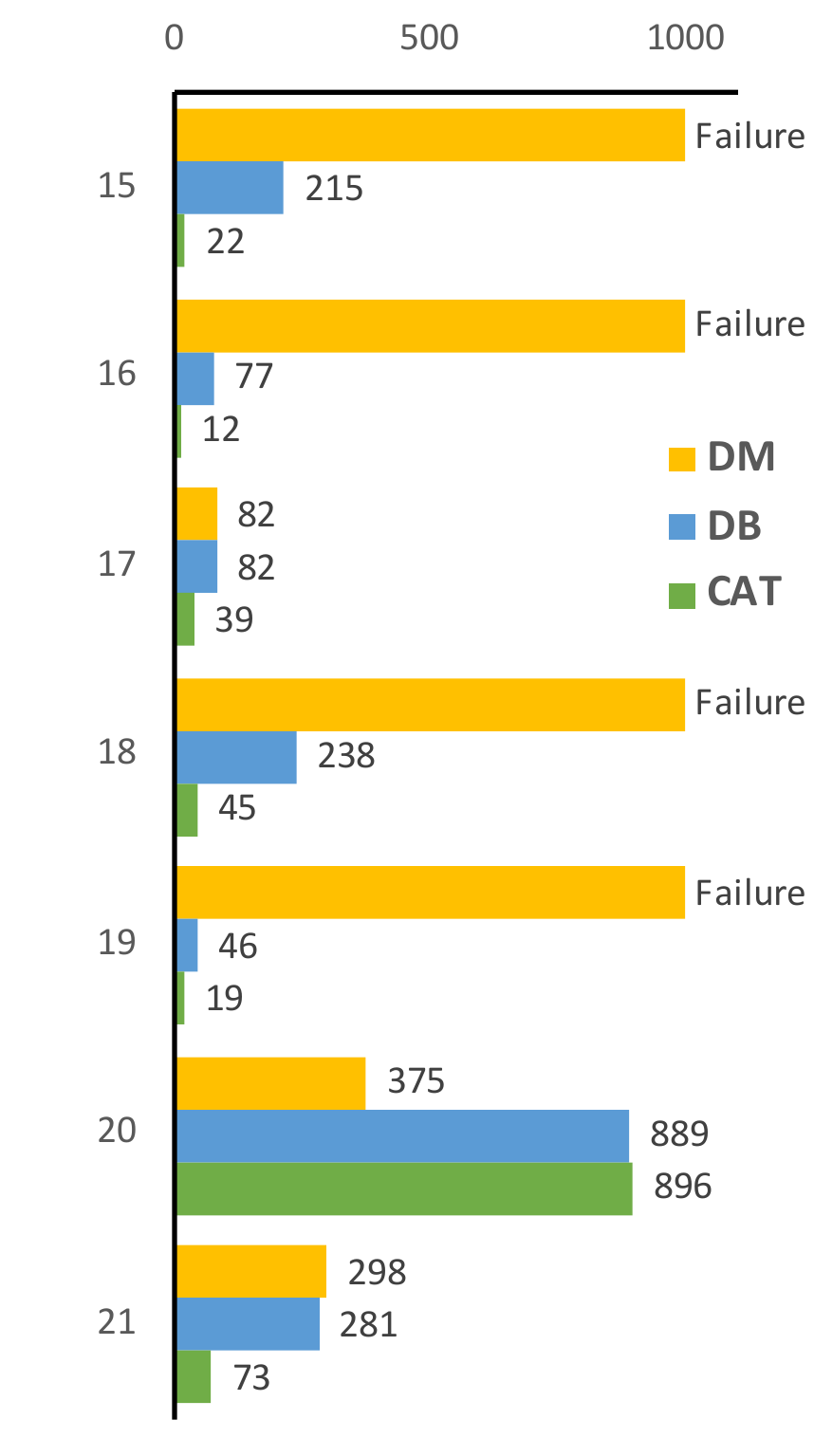}
		
	\end{tabular}
	%\vspace{-12pt}

	\caption{Number of events taken to reach the first target GUI element interaction with DM, DB and CAT.}
		%\vspace{-10pt}
		%\vspace{-12pt}
    \label{fig:tableColumnOne}
\end{figure*}

\begin{figure*}[!htbp]
	\centering
	\includegraphics[trim = 0cm 0cm 0cm 0cm,scale=0.5]{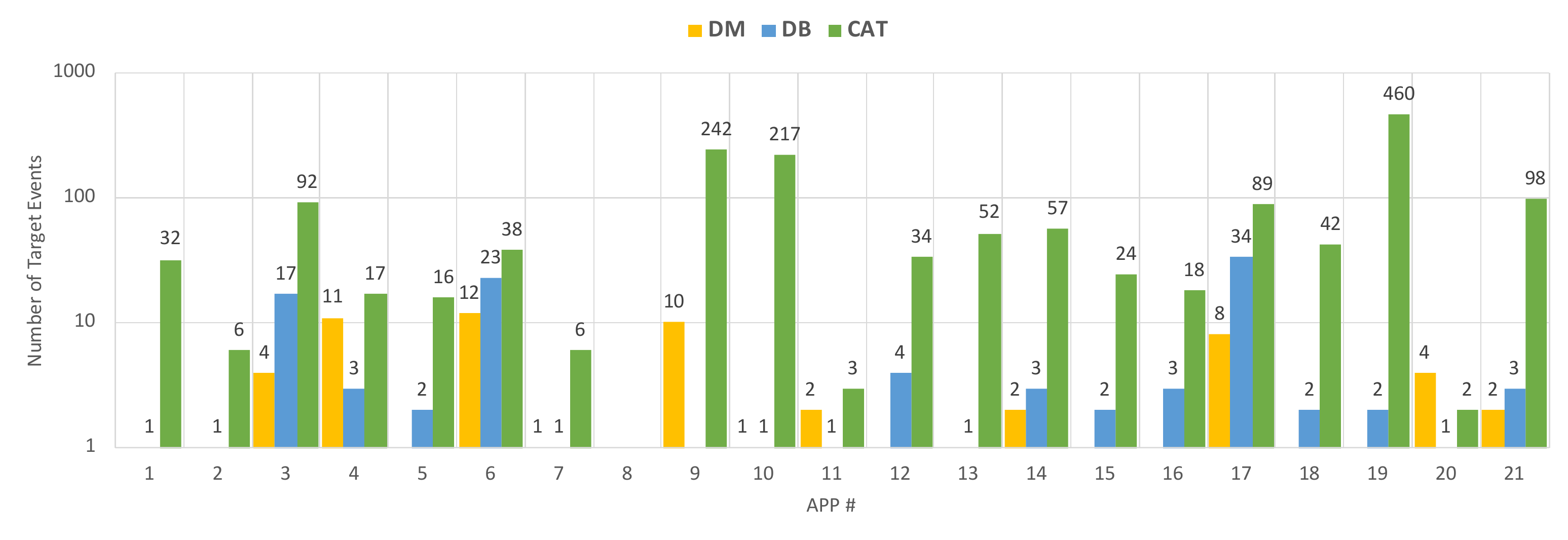}
	\caption{Number of target GUI element interactions achieved by DM, DB and CAT}
	%\vspace{-12pt}

        \label{fig:tableColumnTwo}
\end{figure*}

\paragraph{Number of events to first target element interaction} %We are interested in determining how quickly the tools start interacting with the target GUI element. 
%GUI events that interact with non-target elements is not interesting when testing app changes. 
Figure~\ref{fig:tableColumnOne} shows the average number of events each tool used before it first interacted with the target GUI element for all 21 Apps. Smaller number of events is better, as it indicates the tool starts interacting with the target element faster. %Best performing tool for this metric is emphasied for each app.  
Failure labels on bars indicate the associated tool did not interact with any target GUI element for that App in all 1000 GUI events. 

%When comparing average number of events to first target element interaction, 
%It is evident from Figure~\ref{fig:tableColumnOne} that 
CAT is the fastest to start interacting with the target element, only needing 83 events, on average,  versus 286 for DB and 258 for DM. For 18 of the 21 Apps in Figure~\ref{fig:tableColumnOne}, CAT needs fewer events than DB or DM to start target element interaction. This is because when target state is entered, CAT prioritises interaction with target elements unlike the other two tools. %This feature enable CAT to use fewer events to start interacting with target elements.
%It takes longer when the screens are deeper in the event sequence. 
On average, DM uses 159 events to enter the target state and a further 99 elements to interact with the target GUI element. 
In contrast, depth-first exploration used by DB and CAT enables them to reach deeper screens faster. We remind the reader that CAT diverges from DB only after reaching the target state. Both tools use 81 events, on average, to reach target state. DB uses an additional 205 events to start interacting with the target element. On the other hand, CAT only needs 2 additional events to start target element interaction. %CAT performs better than DB and DM for 19 out of 21 apps. 

DM outperforms DB and CAT on two Apps: \texttt{BeeCount} (App \#3) and \texttt{Trick Tripper} (App \#20). For \texttt{BeeCount}, shown in Figure~\ref{fig:bee-count-example}, the target GUI element is the \texttt{Save} button in the \texttt{New Project} \texttt{Activity}. Entering this \texttt{Activity} requires clicking the \texttt{New Project} button in the start screen. DM clicks this button earlier than DB and CAT (they click on a different button to first go into Settings, perform further events and then return to click the New Project Button). Similarly, For \texttt{Trick Tripper}, the target state containing the target GUI element can be entered after the first screen. DM clicks the button leading to the target state right away while DB and CAT explore many other events and \texttt{Activities} before entering the target state. %  wasted a lot of events trying to create a new trip (new trip creation involves a lot of screens including adding travelers, recording expenses etc.). 
%In this sense, CAT does not waste time and GUI events after it meets the target GUI element and improves the efficiency of test execution.
%In general, CAT outperforms the other two tools for this criteria for 19 out of 21 apps.

\begin{figure}[!htbp]
	\centering
	\includegraphics[trim = 0cm 0cm 0cm 0cm,scale=0.35]{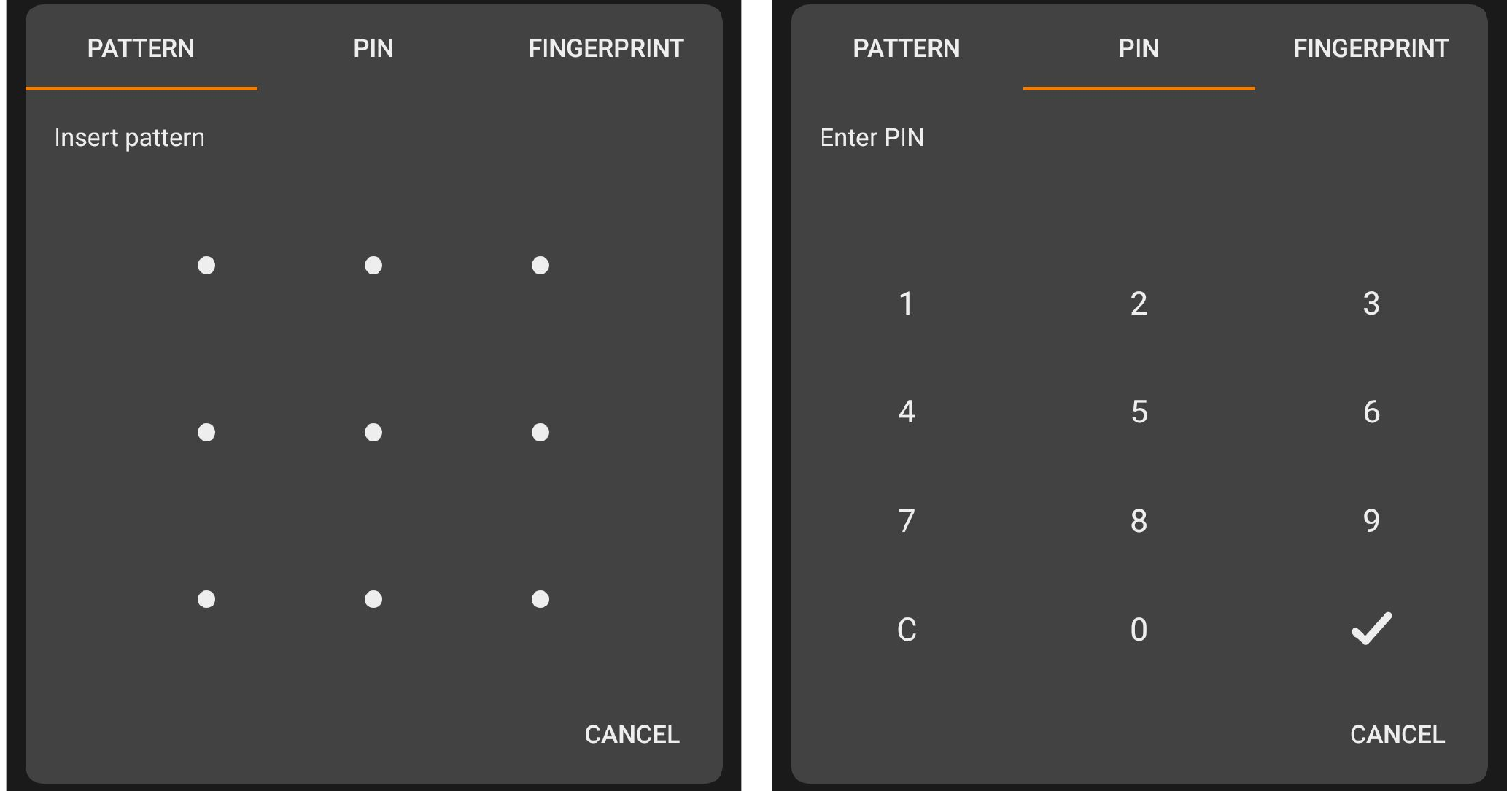}
	%\vspace{-10pt}
    \caption{Setting password in the Simple File Manager App}
   % \vspace{-12pt}
    \label{fig:SimpleFileManagement}
\end{figure}

For \texttt{Simple File Manager} (App \#8), 
all 3 tools failed to interact with the target GUI element within 1000 events. %The 3 tools spent these limited number of events in the Settings activity and read file content activities. 
Target GUI element for this App is a button for creating shortcuts. The state containing this button is not reached by all 3 tools as they all get stuck on a screen for creating passwords, shown in Figure~\ref{fig:SimpleFileManagement}. To be able to leave this state, the tools had to provide input events that were exactly the same for  initial and confirmation password, which none were able to do in 1000 tries.
To avoid getting stuck in such non-target states, CAT provides the developer with an interface to specify an event sequence that leads it directly to the target state.
When such a sequence is provided, CAT is able to interact with the target GUI element 90 times with the first interaction happening after 17 events.

\iffalse
We found that after an average number of 159 events DM generates, it is able to enter the screen containing the target GUI elements. However, it only performed that first interaction with these GUI elements after 258 events. 
According to our observation, we found that DM prefers GUI elements of the first screen it meets. Most of the GUI interactions are those performed at the start screen of these Apps.
By contrast, DB and CAT are able to reach the target state after a mean of 81 events as they utilise a depth-first algorithm. This algorithm enables explorations of as many states as possible in during a limited number of events.
However, DB does not have a preference on which GUI element to interact first and exercise all interactable GUI elements moderately. Therefore, DB starts the first interaction with the target GUI elements after 286 events (106 events after it enters the state containing the target GUI element) while CAT starts exercise the target GUI element after 83 events (right after the target state is entered).
\fi

\paragraph{Number of target element interactions} Figure~\ref{fig:tableColumnTwo} gives the frequency of interactions with the target GUI element. %This measurement gives a sense for how rigorously the the target element is exercised. 
%Ideally, we would like to explore many permutations of target element interaction events and other events (before or after) to see whether the changed impacted GUI element interacts with other GUI elements correctly.
We find, in general,  DM and DB generate few events to exercise target elements. %exercise the target state with several events but generate few events to exercise target elements. 
DM performs 3 target element interactions on average for each App, and DB performs 5 interactions. CAT, on the other hand, performs 69 target element interactions that explore different event orders in length-3 sequences. CAT clearly provides a more rigorous event generation for exercising target GUI elements, outperforming DB and DM on all 21 Apps (as seen in Figure~\ref{fig:tableColumnTwo}). 

\iffalse
%This is because there are several event options in the target state, and they are all given equal priority in DB and DM with no special attention to target elements.  
The average number of events performed by DM when it stayed at the target state is 3. However, DM performs 34 events in total in these states. Similarly, DB performs an average of 5 events of the target GUI element out of a total of 73 events in the target state. By contrast, when CAT enters the target state, it stops exploring more states, resulting in 170 events in the target state (97 more than DB) and 74 events related to the target GUI element (69 more than DB). 
Some events are used after an event in the target state leads to another state to go back to the target state. 
In general, CAT outperforms the other two tools for this criteria for all the 21 subject Apps.
\fi

\paragraph{Coverage of change affected Java code} 

We measured cverage of change impacted Java functions as fraction of change impacted Java functions invoked by the input events. 
Change impact analysis in CAT also gives the set of functions in the code impacted by App updates. 
For Apps in our dataset, we found input events that interacted with the target GUI element invoked all the change impacted functions in code, achieving 100\%  coverage.  This was true for all 3 tools. Results for change affected function coverage is shown in Table~\ref{table:function-coverage}.

\begin{table}[htbp]
    \caption{Change-affected function coverage achieved by DM, DB and CAT}
    \begin{tabular}{|c|l|c|c|c|}
    \hline
    \multirow{2}{*}{\textbf{\#}} & \multirow{2}{*}{\textbf{App Name}} & \multicolumn{3}{c|}{\textbf{Changed-Affected Function Coverage}} \\ \cline{3-5} 
                                 &                                    & \textbf{DM}     & \textbf{DB}     & \textbf{CAT}    \\ \hline
    1                            & World Weather                      & 0\%                    & 100\%                 & 100\%           \\ \hline
    2                            & Amaze File Manager                 & 0\%                    & 100\%                 & 100\%           \\ \hline
    3                            & BeeCount                           & 100\%                  & 100\%                 & 100\%           \\ \hline
    4                            & Diary                              & 100\%                  & 100\%                 & 100\%           \\ \hline
    5                            & Omni Notes                         & 0\%                    & 100\%                 & 100\%           \\ \hline
    6                            & OpenTasks                          & 100\%                  & 100\%                 & 100\%           \\ \hline
    7                            & Simple Draw                        & 100\%                  & 100\%                 & 100\%           \\ \hline
    8                            & Simple File Maganer                & 0\%                    & 0\%                   & 0\%             \\ \hline
    9                            & Simple Solitare                    & 100\%                  & 0\%                   & 100\%           \\ \hline
    10                           & WiFiAnalyzer                       & 100\%                  & 100\%                 & 100\%           \\ \hline
    11                           & Hibi                               & 100\%                  & 100\%                 & 100\%           \\ \hline
    12                           & Geological Timescale               & 0\%                    & 100\%                 & 100\%           \\ \hline
    13                           & DroidShows                         & 0\%                    & 100\%                 & 100\%           \\ \hline
    14                           & Suntimes                           & 100\%                  & 100\%                 & 100\%           \\ \hline
    15                           & Word Scribe                        & 0\%                    & 100\%                 & 100\%           \\ \hline
    16                           & Nani                               & 0\%                    & 100\%                 & 100\%           \\ \hline
    17                           & Fate Sheets                        & 100\%                  & 100\%                 & 100\%           \\ \hline
    18                           & Lift                               & 0\%                    & 100\%                 & 100\%           \\ \hline
    19                           & Currency                           & 0\%                    & 100\%                 & 100\%           \\ \hline
    20                           & Tricky Tripper                     & 100\%                  & 100\%                 & 100\%           \\ \hline
    21                           & Open Money Box                     & 100\%                  & 100\%                 & 100\%           \\ \hline
    \multicolumn{2}{|c|}{Average}                                     & 52\%                   & 90\%                  & 95\%            \\ \hline
    \end{tabular}
    \label{table:function-coverage}
\end{table}
%Thus, tools that exercised the target GUI element in an app non zero times (last column in Table~\ref{table:summary}) achieve 100\% changed function coverage over the app, while  \texttt{failed} values correspond to zero coverage. 
%We do not show code coverage achieved by executing GUI events in Table~\ref{table:summary}. 
%Full function coverage of functions with changes or impacted by changed function is achieved by CAT for apps whose target GUI element is exercised as reported in Table~\ref{table:summary}.

\paragraph{Overhead} DM has the highest overhead in test generation and execution. Average time taken by DM for each App, to generate and execute 1000 events, is 35 minutes while DB and CAT take 18 minutes. CAT introduces a negligible additional overhead of 5 seconds over DB.  %but this is negligible when compared against 18 minutes. 
Higher overhead observed with DM is because DM consumes approximately 1 second more than DB or CAT after every event execution to extract the state model and generate the next event. The extra time accumulated across 1000 events resulted in significant difference. 
%After every event execution and screen state change, DM consumes more time (~1 second) than DB and CAT for model to extract the state model and generate tests, resulting a longer overhead.

\subsection{Q3. Bug Findings}

CAT uncovered changed-related bugs in 2 of the 21 Apps -- \texttt{World Weather} and \texttt{BeeCount}. DM did not reveal bugs in any of the Apps, while DB revealed a bug in the \texttt{World Weather} App but not \texttt{BeeCount}. Both Apps have changes at source code level and change-impacted GUI elements are revealed and tested by CAT. 

Latest version of \texttt{World Weather} changes the function to query weather information using a public API.
CAT identifies the button for changing API key in the \texttt{Settings} \texttt{Activity} as the target GUI element. %which has a link to this function. 
CAT's first interaction with the target GUI element results in a crash and the App stops execution. 
%A crash is observed when this button is clicked and the app stops execution. 
Both DB and CAT are able to crash the App by clicking the button for changing API key. 
However, once the App is restarted, DB does not produce additional events to interact with this target element as it proceeds to interacting with other unvisited elements.  CAT, on the other hand, continues to generate and execute events related to this button as it prioritises target element interaction. 
Interestingly, clicking this button only crashes the first time, further clicks do not result in crashes. This information about first-time only crash is useful for debugging and can only be provided by CAT as it interacts with the target element multiple times. The crash was reported to the App developer and is awaiting a fix at the time of writing this paper. 

\texttt{BeeCount} App version in our dataset updates the listener function for \texttt{Save} button in the \texttt{New Project} \texttt{Activity}. 
CAT identifies the \texttt{New Project} \texttt{Activity} as the target state and the \texttt{Save} button as the target GUI element. The following event sequence generated by CAT -- \texttt{Edit text}, \texttt{Click Menu}, \texttt{Click Save} -- reveals a bug in the App. Clicking the \texttt{Menu} button after editing text in the \texttt{New Project} \texttt{Activity} loses the edited text.  As a result, clicking the \texttt{Save} button does not perform the expected action of saving the edited text. %the menu is shown and text edited disappears after returning from the menu. 
The App is expected to save the text when leaving the New Project \texttt{Activity} so that when the user later returns to this \texttt{Activity} the previously entered text is retained. %Alternatively, the developer could add a confirmation dialog when the user attempts to leave the New Project activity that prompts the user with an option to save the text. 
This bug in the \texttt{BeeCount} App is not revealed by DM and DB. Both DM and DB interact with the target \texttt{Save} button. However, they are unable to generate the event sequence that triggers this bug - Clicking \texttt{Menu} button after editing text, followed by clicking \texttt{Save}. Order of events is important for triggering this bug. %, it is not enough to simply click the \texttt{Save} button. 
CAT's focus on Length 3 event sequences interacting with the target element allows it to trigger bugs that are sensitive to event orders. 

\label{sec:bug-findings-result}

\subsection{Threats to Validity}
\label{sec:threats-to-validity}

A potential threat to internal validity is bugs in CAT's implementation. To mitigate this threat, we conducted careful code reviews and extensive testing. Further, the implementations are publicly available for other researchers and
potential users to check the validity of our results.%We used Java and Python to implement our CAT framework for change impact analysis and test generation respectively. We had a careful code review and made our work publicly available.

A potential threat to the external validity is related to the fact that
the set of Android Apps we have considered in this study may not
be an accurate representation of the App under test. We attempt to reduce the selection
bias by using a dataset of 21 apps from different categories with a variety of Android features.
%For threats to external validity, we use 21 open source Android Apps in the experiment to evaluate our approach. Although they are from different types of application categories and their code covers a variety of features of Android programming, a thorough study on more popular Android applications can further improve the validity of our study. 

A threat to construct validity is caused by restricting the number of GUI events generated by all 3 tools to 1000. Restriction to 1000 input events is used by DM and DB in their default settings and we used the same for CAT. We don't believe changing the number of input events will affect the validity of our results as all 3 tools will be uniformly affected.  
A final threat to validity  is the limited number of tools used in comparison. We used DB and DM as they are popular, well-maintained and easy to use. 
Including other GUI testing tools in the comparison will strengthen the validity of our results. 
%In addition, we use only DM and DB tools to generate test inputs for our subject Apps as they are popular, well-maintained and well-documented. 
%A further comparison with other tools capable of reporting lists of GUI elements their tests interact with and tests written by the developer for the new version of the App can also strengthen the validity of our study.

\section{Conclusion}
\label{sec:conclusion}
We presented the CAT framework for GUI test input generation targeting Android App updates. CAT supports change impact analysis to identify GUI elements affected by updates. %, test input generation and execution focusing on GUI elements impacted by source code changes. 
It then generates length-3 GUI event sequences for interacting with the target GUI elements. 
%CAT takes the APK file of the App under test and changes made to this version as input, builds the GUI element-function graph, computes change impact to the GUI level and generates GUI events as test inputs to exercise the changes.

We empirically evaluated CAT's performance by comparing it to DB and DM over 21 open-source Android Apps. We made the following observations in our experiment.  
%Our empirical evaluation of CAT with 21 open-source Android Apps revealed the following findings,
\begin{enumerate}
    \item CAT is able to trace changes made at the source code level to affected GUI elements automatically. %With both static analysis on the source code and dynamic information gathered at runtime, change-impacted GUI elements and \texttt{Activities} are set as target GUI elements and target states for testing.
    \item For target states containing target GUI elements, CAT is able to generate length-3 event sequences. %Each sequence consists of interactions with the target GUI element and other elements in the state. Different orders of the all the possible sequences are covered in tests it generates.
    \item CAT interacts with target GUI elements sooner than DB and DM, CAT requires 83 events, on average,  versus 286 for DB and 258 for DM. %prioritises entering the target state as early as possible and starting executing generated tests. When comparing to another two popular test case generator for Android, DroidMate and DroidBot, CAT stays at the target state longer and interacts with the target GUI element more frequently than them. 
    \item CAT interacts with the target GUI element more frequently than DB and DM -- average of 69 interactions for CAT, 5 for DB and 3 for DM.
    \item Change-related bugs are revealed by CAT in two Apps. Order of input events was crucial in revealing the bug on one of these Apps. Only CAT was able to reveal this event order sensitive bug owing to the length 3 event sequences used to interact with the target element. %, which emphasises the importance of using various event sequences of interacting with the target GUI element.
\end{enumerate}

In sum, the CAT framework is a novel, automated and effective tool that will help Android developers to test new versions of their Apps focusing on updates. In the future, we plan to optimise CAT's exploration strategy so that target states are entered sooner and the number of events used in other states is reduced.

\bibliographystyle{ieeetr}
\bibliography{references}

\end{document}